\documentclass[peerreview]{IEEEtran}
\usepackage{cite}
\usepackage{graphicx}

\usepackage{algorithm}
\usepackage{algpseudocode}
\usepackage{amsmath}
\usepackage{amsfonts}
\usepackage{amssymb}

\usepackage{hyperref}
\usepackage[margin=1in]{geometry} 
\usepackage{booktabs} 
\usepackage{tabularx} 
\usepackage{listings}
\usepackage{longtable} 
\usepackage{multirow}
\usepackage{adjustbox}

\begin{document}
\title{RPLUW/M: Enabling RPL on the Internet of Underwater Things}
\author{Mohammadhossein Homaei \\
\textit{Departamento de Ingeniería de Sistemas Informáticos y Telemáticos,} \\
\textit{Universidad de Extremadura, Av/ Universidad S/N, 10003, Cáceres, Extremadura, Spain} \\
\href{mailto:Homaei@ieee.org}{Email: Homaei@ieee.org}
}
\date{14 August 2024}

\maketitle

\begin{abstract}
With the widespread use of the Internet of Things, underwater control and monitoring systems for purposes such as ocean data sampling, natural disaster prevention, underwater surveillance, submarine exploration, and the like have become a popular and challenging topic in computers. So far, various topology control and routing solutions have been proposed for these networks. However, as technology expands and applications grow, so does the need for a stable underwater communication platform. On the other hand, underwater communication is associated with challenges such as node mobility, long propagation delays, low bandwidth, limited resources, and high error rates. In this research, for the first time, a topology control platform based on the RPL tree is modelled by applying its structural changes underwater. The proposed RPLUW methods in the case of RPLUWM fixed nodes are introduced to support the mobility of nodes underwater. Flexible objective functions, utilisation of decision-making systems, and application of control schedules in these methods have increased network life, reduced overhead, and increased node efficiency. The simulation results of the proposed method, in comparison with recent methods in this field, show an increase in network efficiency.
\end{abstract}
\vspace{5mm} 
\noindent \textbf{Keywords:} Internet of Underwater Things, Routing, RPL tree, Mobility, Decision Systems, Network Lifetime.

\section{Introduction}\label{sec1}
With the increasing use of Internet of Things (IoT) applications, underwater acoustic sensor networks have become an essential part of this technology in marine science for researchers and marine-related industries. Nowadays, with the integration of telecommunication and computing platforms, this issue is also recognised as part of the more comprehensive underwater IoT problem for stationary and mobile sensors/actuators and underwater robots. Underwater monitoring in the seas and oceans is vital due to their different military, environmental, and industrial applications \cite{Mary2021_2}. Previously, underwater communications focused mainly on physical layer communications and signal processing issues, and there was little networking discussion. Since applications such as underwater environment monitoring are performed on a large scale, expanding the underwater network is inevitable, \cite{Luo2021_4}. Many of these sensors (sonar, optical instruments, laser, magnetic, etc.) are placed underwater to carry out the monitoring process. Expanding the monitoring environment requires properly analysing sensor output and networking \cite{Bello2022_10}. Most underwater ecosystems are high-risk environments; therefore, the limited resources of the underwater sensor network require performance reliability and stability more than a conventional sensor network.

Routing protocols in computer and telecommunication networks are essential to network performance \cite{Ismail2022_11}. Proposed protocols for underwater sensor networks can be divided into two general parts: location-based and location-independent routing protocols. We know that water currents and sea creatures move randomly; therefore, location-based routing protocols are unsuitable for underwater environments. On the other hand, using the GPS global positioning system in an underwater environment is inadequate. In underwater wireless sensor networks, the nodes are often battery-powered, and it is impossible to recharge the battery; therefore, routing protocols must be optimised regarding power consumption to communicate between sensor nodes. These protocols must be able to store energy and consume it reasonably in exchange for error-free communication and data transmission \cite{Chaudhary2023_13, Mohsan2023_12}.
On the other hand, when the sensors collect the required information, they must send it to the water level's base station. Transmitting information from sensor nodes to the base station is very expensive in terms of energy consumption; hence, energy consumption is one of the vital factors in designing routing protocols for underwater wireless sensor networks. There are many limitations to the underwater environment, and some of the most critical issues to consider when designing underwater sensor network routing protocols and the IoUT platform include the following \cite{Jiang2023_14, tarif2023review}:
\begin{itemize}
    \item Energy issue: Energy is a limitation of the underwater sensor network, as batteries do not have solar energy to charge and are not easily replaced. Therefore, routing protocols should consider energy saving as a key element because the node is dead after the energy runs out and may cause the project to fail.
    \item Load Balancing: An optimal routing protocol uses network resources fairly and equitably. This approach can prevent the occurrence of Bottlenecks or Hotspots. Also, in case of such incidents, action should be taken to resolve the issue as soon as possible \cite{Bello2022_10}.
    \item Underwater Location: One of the important features that all routing protocols in the underwater network suffer from is the lack of GPS location information for nodes and their neighbours \cite{Nkenyereye2024_16, Ismail2022_11}. Knowing the location of nodes and neighbours reduces routing tables, reduces neighbour-finding efforts, and prevents loops and wandering packages in the network. Unfortunately, GPS was associated with errors in shallow water (less than 4 meters) and is unusable in deep water \cite{Khan2022_7, Ali2023_15}. In the underwater network, the nodes use only one Z-axis, known as the depth gauge, and are equipped with it. A routing protocol with only the depth of the node from the water level and the neighbour’s list should act to transfer the node information to the base station at the water level, which is an unresolved challenge.
    \item Node mobility and instability of the fluid environment: The instability of the nodes underwater due to environmental factors such as hot and cold water currents, collisions of underwater organisms, and fluid waves is apparent \cite{Shovon2022_8}. Since the sensor nodes are connected to the seabed by chains and are suspended between the seabed and the sea surface, they are constantly moving. The node's mobility causes the list of neighbours, the path is chosen, and the path to be explored and repaired to undergo fundamental changes.
    \item Lack of a fault detection system: If a failure or underwater network configuration problem occurs, it is not detected before retrieving and aggregating network data. This process may easily lead to the complete failure of the monitoring mission \cite{Ali2023_15}.
    \item Lack of real-time monitoring: Recorded data are unavailable at the base station until collection and processing. This process may occur several hours after each sampling \cite{Chaudhary2023_13}.
    \item Impossibility of instantaneous system configuration: Interaction between coastal control systems and monitoring commands is impossible in real-time. This prevents the adaptive set of commands, and it is impossible to configure the system after a specific event \cite{Nkenyereye2024_16}.
\end{itemize}
According to the issues as mentioned earlier, there is a need to develop underwater network protocols for real-time monitoring of ocean basins. Due to the many environmental changes in the underwater environment, efficient protocols must be designed to meet the needs of this stressful environment \cite{Jiang2023_14, Ismail2022_11}. The specific features of the underwater communication channel, such as limited bandwidth capacity and long delays, require new, reliable, and efficient data communication protocols. Therefore, in this paper, for the first time, we have adapted the RPL protocol developed in recent years by the IETF Group for the Internet of Things and LLN networks on the ground to the underwater environment \cite{Richardson2021_17}. For this purpose, our changes have been in the physical layer, the data layer, the network layer, and the transmission layer. Also, by adding the mobility of nodes in this protocol, it is possible to make simulations that are very close to reality.
 
The second part of this article is divided as follows: Section \ref{sec2} provides an overview of routing methods and algorithms in the underwater sensor network. Section \ref{sec3} introduces an optimized RPLUW routing protocol with mobility support capability, addressing some of the challenges and limitations discussed in Section \ref{sec2}. In Section \ref{sec4}, we detail the system model and thoroughly evaluate the proposed method, comparing it to other mixed methods. Section \ref{sec5} focuses on the simulation results, demonstrating the effectiveness of the proposed approach. Finally, Section \ref{sec6} concludes the article by summarising findings and suggesting future research directions.

\begin{table*}[ht!]
\caption{Abbreviations and Acronyms/ Variables and Definitions}
\label{table:variables_abbreviations}
\centering
\begin{tabular}{|p{0.07\textwidth}|p{0.3\textwidth}|p{0.07\textwidth}|p{0.3\textwidth}|}
\hline
\textbf{Variable} & \textbf{Definition} & \textbf{Variable} & \textbf{Definition} \\
\hline
u, v & Node in graph & DPP & DODAG Preferred Parent \\
T & Temperature in degrees Celsius & DRL & DODAG Root List \\
S & Set of nodes & d & Depth in meters \\
B & Border routers & f & Frequency \\
N & Noise & $\bar{k}$ & Constant K is degree of graph \\
p & Parent node & p' & Alternative parent node \\
DODAG & Destination Oriented Directed Acyclic Graph & Root & Root node of graph (Sink) \\
$\varepsilon$ & Shipping factor (0–1) & DIO & DAG Information Object \\
w & Wind speed & DAO & Destination Advertisement Object \\
P & Power of signal & DIS & DODAG Information Solicitation \\
CC & Capacity of channel & DAO-Ack & Acknowledgement for DAO \\
DT & Delay time & IC & Inconsistency \\
C & Consistency & I$_{\text{min}}$ & Minimum interval \\
DT$_{\text{proc}}$ & Processing delay time & I$_{\text{max}}$ & Maximum interval \\
DT$_{\text{queue}}$ & Queuing delay time & ND & Neighbour Discovery \\
DT$_{\text{trans}}$ & Transition delay time & NS & Neighbour Solicitation \\
$\rho$ & Constant factor & NA & Neighbour Advertisement \\
$\alpha$ & Absorption & RS & Router Solicitation \\
T$_{\text{rx}}$ & Transition time & RA & Router Advertisement \\
r & Radios & L$_{\text{t}}$ & Linkage Timer \\
R & Number hops of Root node & M$_{\text{t}}$ & Mobility timer \\
J & Root node in graph & R$_{\text{t}}$ & Response Timer \\
OF & Objective Function & M & Alive nodes \\
G & Graph & \(\tau\) & Predetermined lifetime \\
V & Set of vertices & E & Set of edges \\
\hline
\end{tabular}
\end{table*}

\section{Related works} \label{sec2}
\subsection{Underwater sensor routing without the location information}
The purpose of creating an underwater sensor network is to sample the physical parameters of the environment and form a graph to transfer data to the central node. Therefore, several methods have been introduced to manage data transfer. It is impossible to know the location of network nodes in the underwater environment due to the lack of GPS signals, and the location of each node is constantly changing due to environmental factors. In Table \ref{tab:comparison-protocols-part1}, we have examined models that operate on graph formation, routing, and data transmission without knowing the node's location.

\begin{table*}[htbp]
\caption{QoS-aware Underwater IoT Routing Protocols (Part 1)}
\label{tab:comparison-protocols-part1}
\centering
\resizebox{\textwidth}{!}{%
\begin{tabular}{|p{1cm}|p{6cm}|p{4.5cm}|p{3.5cm}|}
\hline
\textbf{Year/Ref} & \textbf{Aim/Strategy} & \textbf{Strengths} & \textbf{Parameters} \\
\hline
2020 \cite{V2020_21} & Clustered geographic-opportunistic routing protocol (C-GCo-DRAR) for UWSNs. It aims to address challenges like high propagation delay and energy constraints through clustering and depth-based topology adjustments. & Demonstrated superior performance in packet delivery, energy efficiency (EE), and reduced delays via Aquasim simulator. Utilizes energy levels for cluster head election and depth adjustment for void recovery. & Packet Delivery Ratio(PDR), EC, E2E Dealy. \\ 

\hline
2020 \cite{Usman2020_22} & Energy-efficient routing in IoT-enabled UWSNs for smart cities using "Underwater (ACH)²" (U-(ACH)²). Incorporates depth considerations to optimize energy use across varied deployment scenarios. & Outperforms DBR and EEDBR in packet delivery rates, energy usage, and network lifetime (NL), promising for smart city applications. & EC, PDR, NL. \\ 

\hline

2020 \cite{Wang2020_23} & Game-Theoretic Routing Protocol (GTRP) for 3-D Underwater Acoustic Sensor Networks (UASNs). Utilizes a strategic game with Nash equilibrium for packet forwarding, minimizing broadcasts for node degree estimation. & Shows enhanced packet delivery, reduced delay, and EE in Aqua-Sim simulations. Addresses latency, mobility, and bandwidth challenges effectively. & PDR, E2E delay, EC. \\ 

\hline
2020 \cite{Li2020_24} & Distributed Multiagent Reinforcement Learning (DMARL) protocol for Underwater Optical Wireless Sensor Networks (UOWSNs). Focuses on dynamic topologies and energy optimization through distributed decision-making. & Improved energy usage, PDR, and load distribution validated through simulations. Demonstrates adaptability and efficiency in UOWSNs. & EC, PDR, load distribution. \\ 

\hline
2020 \cite{Natesan2020_25} & Fuzzy Logic Cluster-Based Energy Efficient Routing Protocol (FLCEER) for UASNs. Implements multi-layer clustering and fuzzy logic for efficient routing and UCH election. & Enhances EE, packet delivery, throughput, and NL, outperforming MLCEE, DBR, and EEDBR in simulations. & EE, PDR, throughput, nNL. \\ 

\hline
2020 \cite{Hussain2020_26} & Sleep-Scheduling Oil Detection Routing Protocol for UWSNs in smart oceans. Integrates IoT for energy-efficient oil spill detection using a 2D network architecture and sleep scheduling. & Extends NL and improves detection efficiency, focusing on environmental monitoring and protection. Demonstrates energy conservation in simulations. & Energy conservation, detection efficiency, NL. \\ 
\hline

2020 \cite{Faheem2020_27} & FFRP introduces a self-learning dynamic firefly mating optimization for efficient and reliable data routing in IoUT. & Superior packet delivery ratio, lower latency and EC, enhancing network throughput. & Potential complexity in real-world deployment due to the bio-inspired, computation-intensive optimization process. \\ 
\hline

2020 \cite{Coutinho2020_28} & Stochastic modelling of opportunistic routing for IoUT, leveraging programmable physical layers and multi-modal communication. & Improved data delivery rates through innovative candidate-set selection, integrating acoustic modem and node selection. & Increased Energy Consumption trade-off, requiring efficient energy management strategies for practical application. \\ 
\hline

2020 \cite{Robinson2020_29} & Hybrid optimization routing for AUVs in IoUT, focusing on EE and effective data collection via A-ANTD and TARD phases. & Reduces energy usage, improves data delivery efficiency, and enhances network performance for smart ocean applications. & The complexity of coordinating AUVs and sensor nodes might limit scalability and adaptability in diverse aquatic environments. \\ 
\hline

2020 \cite{Coutinho2020_31} & A novel Power-Controlled Routing (PCR) protocol for IoUT that dynamically adjusts transmission power based on environmental conditions. & Improves energy use and data delivery rates through dynamic power control and opportunistic routing. & Complex adjustment algorithms may increase computational overhead. \\ 
\hline

2021 \cite{Arul2021_32} & Utilizes Intelligent Data Analytics (IDA) for Optimized Energy Planning (OEP) in IoUT, enhancing data transmission efficiency and energy optimization. & Significant increase in packet delivery rate and latency and energy expenditure reductions. & The dual-stage programming framework could be complex to implement and manage in real-time. \\ 
\hline

2021 \cite{Majid2021_33} & Discusses green energy harvesting and energy-efficient routing for IoUT, exploring sustainable and renewable energy sources. & Focuses on sustainability and tapping into unexplored energy resources, potentially reducing dependency on traditional power sources. & May require substantial initial investment and infrastructure for energy harvesting technologies. \\ 
\hline

2021 \cite{Wang2021_34} & Introduces an Adaptive-Location-Based Routing Protocol (UA-RPL) for UASNs, focusing on optimizing packet forwarding in three-dimensional spaces. & Enhanced network throughput and PDRs, reduced EC and communication delays. & The protocol's efficiency could diminish in extremely dense or highly dynamic underwater environments. \\ 
\hline

2021 \cite{Kumar2021_35} & Examines demur and routing protocols in UWSNs for IoUT applications, aiming to support smart city initiatives. & Highlights the potential of IoUT in environmental monitoring, underwater exploration, and disaster prevention. & Specific challenges and disadvantages related to the implementation of smart cities are not detailed. \\ 
\hline

\end{tabular}
}
\end{table*}

\newpage

\begin{table*}[htbp]
\caption{QoS-aware Underwater IoT Routing Protocols (Part 2)}
\label{tab:comparison-protocols-part2}
\centering
\resizebox{\textwidth}{!}{%
\begin{tabular}{|p{1cm}|p{6cm}|p{4.5cm}|p{3.5cm}|}
\hline
\textbf{Year/Ref} & \textbf{Aim/Strategy} & \textbf{Strengths} & \textbf{Parameters} \\
\hline

2021 \cite{Kapileswar2021_36} & Energy-efficient routing in IoT-based UWSN using the Bald Eagle Search (BES) algorithm. Emulates bald eagle hunting behaviour for optimizing routing, comprising initialization, construction, and data transmission phases for effective energy use and path efficiency. & Demonstrates superior performance in EC, average residual energy, and NL over existing algorithms, addressing critical issues of E2E delay, EC, and reliable data delivery. & EC, average residual energy, NL, PDR, E2E (E2E) delay. \\ 

\hline
2021 \cite{Zhang2021_37} & To enhance reliability, reduce delay, and improve EE in UASN using RLOR. / Merges opportunistic routing with reinforcement learning for dynamic node selection. & Demonstrated superior performance in reliability, low delay, and EE. Innovative recovery mechanism for void navigation. & Complexity of implementing reinforcement learning algorithms in real-time underwater environments. \\ 
\hline

2021 \cite{Khan2021_38} & Address energy consumption and void avoidance in UASNs with QL-EEVARP. / Uses Q-learning for dynamic, energy-efficient path selection and void avoidance. & Achieves better PDR and enhanced EE; adaptive void recovery enhances network performance. & Scalability issues in larger networks; the complexity of Q-learning algorithm implementation. \\ 
\hline

2021 \cite{Khasawneh2021_39} & Improve IoUT data dissemination by mitigating void zones. / EVA framework focuses on preemptive void identification and intelligent routing. & Reduced energy consumption, extended NL, improved packet delivery, and reduced latency. & Advanced algorithms for void detection and navigation may increase system complexity. \\ 
\hline

2021 \cite{Draz2021_40} & Optimize underwater communication in IoUT with dynamic path adjustment. / ROBINA uses Path-Adjustment and path-loss models to maintain data flow in aquatic conditions. & Improved packet transmission, reduced transmission, and path loss; adaptively managed underwater routing. & Deployment complexity in variable environments due to intricate path adjustment mechanisms. \\ 
\hline

2021 \cite{Draz2021)41} & Facilitate reliable and energy-efficient UIoT communication. / ELW-CFR employs proactive routing with layering and watchman nodes for collision-free communication. & Low E2E delay and high PDR; address void hole challenge effectively. & The layering model and watchman node reliance may complicate implementation. \\ 
\hline

2022 \cite{Narla2022_42} &  Enhance energy efficiency in UWSNs through optimized power control. / Introduces a power-controlled routing protocol that dynamically adjusts TPL based on various factors. & Significant improvements in data delivery rates and network longevity; optimizes energy usage. & Potential complexity in dynamic power adjustment and monitoring for effective implementation. \\ 
\hline

2022 \cite{Kulla2022_43} & Enhance communication efficiency in underwater IoT with FBR./ Evaluate FBR performance across different angles to optimize resource use and packet delivery. & Narrower FBR angles led to better performance metrics, including energy conservation and reduced buffer strain. & Configuration of optimal FBR angles is critical and may not fit all operational scenarios in underwater IoT. \\ 
\hline

2022 \cite{Subramani2022_44} &  Optimize energy efficiency in UWSNs. /Metaheuristic-based clustering with Routing Protocol employing CEPOC for clustering and MHR-GOA for routing. & Notable improvements in energy efficiency and network lifespan; effective load balancing in data transmission. & Complex algorithm integration may challenge real-time applicability and scalability in diverse underwater conditions. \\ 
\hline

2022 \cite{Shen2022_45} & Enhance underwater IoUT communication. / Cooperative Routing Protocol based on Q-Learning for hybrid optical-acoustic networks, optimizing connectivity and energy use. & Improved network connectivity, lifetime, and efficiency; reduced packet loss and E2E delay. & Deployment complexity due to hybrid optical-acoustic communication needs and the learning-based routing decision process. \\ 
\hline

2022 \cite{Hu2022_46} & Improve UWSNs' energy efficiency and network longevity. / Cooperative-Communication Based Underwater Layered Routing, integrating cooperative communication with hierarchical clustering. & Extended NL, improved throughput and packet delivery; effective energy consumption balance. & The intricate clustering and cooperative communication mechanisms may complicate protocol deployment. \\ 
\hline

2022 \cite{Gavali2022_47} & Enhance energy efficiency and data transmission in UWSNs. / Energy Optimization using Swarm Intelligence (EORO) protocol, employing EFF-PSO for optimal forwarder node selection. & Superior throughput, EC, and latency metrics; improved PDR. & Complexity of swarm intelligence algorithms might increase computational overhead and affect real-time performance. \\ 
\hline

2022 \cite{Nellore2022_48}  & Mitigate signal transmission challenges in UWSNs. / Utilizes IoT and SNR analysis with OSDM modulation and improved channel estimation for efficient signal transmission. & Enhanced communication efficiency with improved SNR, reduced BER, and minimized MSE. & The complexity of advanced modulation techniques and channel estimation may limit adaptability to all underwater conditions. \\ 
\hline

2022 \cite{Dogra2022_49} & Extend network longevity and improve IoT WSN connectivity. / ESEERP optimizes CH selection using a Sail Fish Optimizer (SFO) for efficient route selection. & Achieves significant improvements in network longevity, energy utilization, and PDR. & The optimization technique's complexity could impact the protocol's scalability and adaptability to varying network sizes. \\ 
\hline

2022 \cite{Celik2022_50} & Optimize underwater IoUT communication. / Sector-based opportunistic routing (SectOR) integrates optical and acoustic communications to enhance packet delivery. & Significant improvements in underwater networks' EE, delay reduction, and PDR. & Challenges in balancing communication range and beamwidth for optimal performance across underwater environments. \\ 
\hline
\end{tabular}
}
\end{table*}

\newpage
\begin{table*}[htbp]
\caption{QoS-aware Underwater IoT Routing Protocols (Part 3)}
\label{tab:comparison-protocols-part3}
\centering
\resizebox{\textwidth}{!}{%
\begin{tabular}{|p{1cm}|p{6cm}|p{4.5cm}|p{3.5cm}|}
\hline
\textbf{Year/Ref} & \textbf{Aim/Strategy} & \textbf{Strengths} & \textbf{Parameters} \\
\hline

2022 \cite{Karim2022_51} &  Evaluate the efficacy of various IoUT routing protocols. / Simulation-based analysis of cluster-based and chain-based routing protocols to enhance efficient data transfer in UWSNs. & Comprehensive comparison revealed cluster-based protocols show varied efficiency, offering insights into effective routing strategies in IoUT. & Requires extensive simulations to capture real-world complexities and underwater conditions accurately. \\ 
\hline

2022 \cite{Sathish2022_52} &  Optimize data transfer performance in UWSNs.  /Perform a performance analysis of diverse routing protocols like AODV, DSR, and DYMO under varying conditions using the QualNet simulator. & Identified protocols with lower power consumption and higher energy efficiency, which are crucial for improving UWSN performance. & Simulation-based approach may not fully replicate underwater environments' unique physical and chemical challenges. \\ 
\hline

2022 \cite{ShakerReddy2022_53}&  Enhance energy efficiency and information transmission in IoT-UWSNs. / Introduces adaptable power networking methods using Fastest Route Fist (FRF) and a business unit method for effective routing. & Proposed methods significantly reduce Electric Cost (EC) and Transmission Drop Rates (RTDR) with reasonable latency. & Complexity of implementing and tuning the proposed adaptable power networking methods in diverse underwater scenarios. \\ 
\hline

2022 \cite{Sun2022_55} & Enhance routing efficiency and energy conservation in UWSNs. / Introduces the Adaptive Clustering Routing Protocol (ACRP) with multi-agent reinforcement learning for adaptive cluster head selection, reducing communication overhead and EC. & Demonstrated improved routing efficiency, energy utilization, and network lifespan compared to existing methods. Efficiently mitigates hotspot issues through balanced EC. & Implementation complexity due to reinforcement learning integration requires rigorous tuning to effectively adapt to diverse underwater environments. \\ 
\hline

2022 \cite{Sathish2022_56}&  Analyze UWSN performance using diverse routing protocols.  /Evaluate protocols like AODV and DSR using simulations to explore their efficacy under various network conditions. & Provided comparative insights into protocol performance, identifying those with potential for UWSN enhancements. & Simulation-based evaluations may not fully capture the operational complexities of real-world underwater environments. \\ 
\hline

2022 \cite{AlBzoor2022_57} &  Enhance IoUT communication efficiency with DSPR.  / Utilizes angle of arrival and depth information for directional data forwarding and selective power control to optimize energy use. & Demonstrated energy efficiency, achieving better performance in delivery ratios and network longevity. & May require sophisticated hardware to accurately determine the angle of arrival and implement selective power control effectively. \\ 
\hline

2022 \cite{KesariMary2022_58} &  Review energy optimization techniques in UIoT.  / Evaluates various energy optimization strategies, including wireless power transfer and artificial intelligence, to enhance network efficiency. & Highlighted potential efficiencies from mixed-medium communication and smarter battery management, identifying research gaps and future directions. & The breadth of the review may necessitate further empirical testing to validate the effectiveness of proposed optimizations in real-world applications. \\ 
\hline

2022 \cite{Lilhore2022_59}&  Address energy optimization in UWSNs.  / Proposes an energy-efficient routing protocol leveraging genetic algorithms for optimal routing and data fusion techniques for energy conservation. & Showed improvements in PDR and EC, offering a viable solution for extending NL. & The complexity of the genetic algorithm and data fusion process may impact the scalability and real-time applicability of the protocol. \\ 
\hline

2022 \cite{Alghamdi2022_60} & Adaptive Transmission-based Geographic and Opportunistic Routing (ATGOR) protocol for UIoTs. Introduces a two-part strategy focusing on cube selection for transmission reduction and reliable node selection for optimal data forwarding. Incorporates Mobility Aware ATGOR (MA-ATGOR) to predict neighbour locations to avoid voids and ensure packet delivery. & Enhances packet delivery reliability, reduces void nodes, and optimizes energy consumption per packet in harsh underwater environments. & PDR, the number of void nodes, and EC per packet. \\ 

\hline
2022 \cite{Fang2022_61} & Stochastic Optimization-Aided Energy-Efficient Information Collection for IoUT. Utilizes heterogeneous AUVs for data collection, optimizing energy efficiency with Particle Swarm Optimization (PSO) and Lyapunov optimization considering AUV trajectory, resource allocation, and Age of Information (AoI). & Offers a holistic approach to optimizing energy usage and AoI in IoUT networks. Successfully balances energy consumption with system stability and information freshness through adaptive planning and optimization strategies. & EC, queue lengths, Age of Information (AoI). \\ 

\hline
2022 \cite{Liu2022_62} & Energy-Efficient Guiding-Network-Based Routing (EEGNBR) for UWSNs. Establishes a guiding network to direct packets via the shortest route with minimal hops, incorporating a concurrent working mechanism for reduced forwarding delay and energy conservation. & Reduces network delay significantly while ensuring reliable routing and EE. Innovative use of guiding network and concurrent data forwarding mechanism. & Network delay, EC, PDR, network service life. \\ 

\hline
2022 \cite{Saravanan2022_63} & Underwater Adaptive RPL (UA-RPL) for IoUT. Modifies RPL's Objective Function (OF) and DODAG construction to improve NL and reliability in underwater conditions. Introduces dynamic trickle algorithm to reduce control message overhead. & Enhances communication reliability and EE in underwater IoUT networks. Successfully mitigates the impact of underwater noise and balances energy consumption across nodes. & PDR, throughput, control overhead, delay, EC. \\ 
\hline
\end{tabular}
}
\end{table*}

\newpage
\begin{table*}[htbp]
\caption{QoS-aware Underwater IoT Routing Protocols (Part 4)}
\label{tab:comparison-protocols-part4}
\centering
\resizebox{\textwidth}{!}{%
\begin{tabular}{|p{1cm}|p{6cm}|p{5cm}|p{3cm}|}
\hline
\textbf{Year/Ref} & \textbf{Aim/Strategy} & \textbf{Strengths} & \textbf{Parameters} \\
\hline

2023 \cite{Guo2023_64} & Opportunity Routing protocol based on Density Peaks Clustering (ORDP) for IoUT. Utilizes network clustering with Density Peaks Clustering (DPC), entropy weight-TOPSIS for cluster head election, and opportunistic data transmission. & Innovatively combines DPC with entropy weight-TOPSIS for efficient cluster head selection, significantly improving EE, transmission latency, and PDR. & EC, average transmission latency, PDR. \\ 

\hline
2023 \cite{Ullah2023_65} & Delay and Reliability Aware Routing (DRAR) and Cooperative DRAR (Co-DRAR) protocols for UWSNs. Aims to enhance reliability with strategies for reducing delay and managing power consumption through regional network division and strategic sink node positioning. & Introduces cooperative transmission to improve data packet quality, effectively reducing E2E delay, balancing energy consumption, and ensuring reliable communication. & EC, E2E delay, PDR, dead nodes, packet drop ratio, alive nodes. \\ 

\hline
2023 \cite{Shah2023_66} & Neighboring-Based Energy-Efficient Routing Protocol (NBEER) for UWSNs. It focuses on Neighbor Head Node Selection (NHNS), cooperative load balancing, and performance enhancement mechanisms. & Excels in reducing energy consumption and latency while improving packet delivery ratio, NL, and total received packets through efficient neighbor-based routing and data forwarding. & EC, E2E delay, PDR, alive nodes, number of packets received. \\ 

\hline
2023 \cite{Sazzad2023_67} & Designing an Underwater-Internet of Things (U-IoT) network model for marine life monitoring. Utilizes autonomous underwater vehicles (AUVs) and surface sinks for efficient data transfer using acoustic waves and RF techniques. & Addresses the overfishing problem by providing a system that supports effective marine life monitoring and data management, demonstrating efficient administration through the proposed network model. & Efficiency of data transfer, management of marine resources, impact on overfishing. \\ 

\hline
2023 \cite{Zhu2023_68} & Shared Underwater Acoustic Communication Layer Scheme (SUACL) for enhancing UAC technology development and evaluation. Enables remote operation of communication units for data transmission and reception. & Offers a flexible and adaptable platform for underwater acoustic research, significantly improving communication efficiency with better SNR, lower BER, and minimized MSE. & Signal to Noise Ratio (SNR), Bit Error Rate (BER), Mean Square Error (MSE). \\ 

\hline
2023 \cite{Jiang2023_69} & Opportunistic Hybrid Routing Protocol (RAOH) for Acoustic-Radio Cooperative Networks (ARCNet). Introduces a hybrid routing strategy that utilizes surface radio links for neighbor discovery and combines opportunistic and on-demand routing for efficient data forwarding. & Enhances packet delivery success, reduces route establishment times, and improves EE by leveraging the dual advantages of acoustic and radio communication. & Energy usage, average transmission latency, PDRs. \\ 

\hline
2023 \cite{Simon2023_70} & Opportunistic Routing-Based Reliable Transmission Protocol (OR-RTP) utilizing Artificial Rabbits Optimization (ARO) for energy-efficient routing in UIoT networks. Focuses on balancing energy consumption and PDR through meta-heuristic relay selection. & Offers an adaptive relay selection mechanism for dynamic underwater environments, improving network longevity and reliability while reducing overall energy consumption. & EC, PDR, throughput, NL. \\ 

\hline
2023 \cite{Jiang2023_71} & Opportunistic Routing based on Directional Transmission (ORDT) for IoUT. Utilizes directional transmission for energy focus, improving packet delivery rates, minimizing latency, and conserving energy. & Combines directional transmission with opportunistic routing for targeted energy use and enhanced packet delivery, addressing underwater communication's unique challenges. & Packet delivery success rate, transmission latency, energy usage. \\ 

\hline
2023 \cite{Su2023_72} & Hybrid-Coding-Aware Routing Protocol (HCAR) for UASNs. Introduces interflow network coding within a reactive and opportunistic routing framework to enhance packet transmission efficiency and network performance. & Integrates network coding to correct errors and optimize transmission efficiency, significantly reducing transmission counts and adapting to UASN conditions. & EC, PDR, throughput, NL. \\ 

\hline
2023 \cite{Sathish2023_73} & Member Nodes Supported Cluster-Based Routing Protocol (MNS-CBRP) for UWSNs. Utilizes clustering and leverages network member nodes for efficient information transfer, optimizing energy consumption. & Improves scalability and data integrity through clustering, significantly extending the network's lifespan by optimizing energy use and enhancing data transmission reliability. & EC, PDR, throughput, NL, energy trade-off. \\ 

\hline
2023 \cite{Guo2023_74} & Efficient Geo-Routing-Aware MAC Protocol (GO-MAC) based on OFDM for UANs. Integrates geo-routing with OFDM, optimizing transmission delay and energy consumption through a cross-layer MAC protocol. & Reduces data collisions and enhances EE with optimized OFDM resource allocation and improved next-hop selection. & EC, PDR, throughput, NL. \\ 

\hline
2023 \cite{Haseeb2023_75} & Energy-Depth Aware Channel Routing Protocol (ED-CARP) for UWSNs in IoUT. It focuses on relay node selection based on Channel State Information (CSI), considering residual energy and depth. & Combines energy and depth awareness in relay selection, optimizing energy consumption and enhancing data delivery efficiency. & EC, PDR, throughput, NL, and balance between energy used in transmission and reception. \\ 

\hline
\end{tabular}
}
\end{table*}

\newpage
\begin{table*}
\caption{QoS-aware Underwater IoT Routing Protocols (Part 5)}
\label{tab:comparison-protocols-part5}
\centering
\resizebox{\textwidth}{!}{%
\begin{tabular}{|p{1cm}|p{6cm}|p{4.5cm}|p{3.5cm}|}
\hline
\textbf{Year/Ref} & \textbf{Aim/Strategy} & \textbf{Strengths} & \textbf{Parameters} \\
\hline

2024 \cite{Ahmad2024_76} & Machine Learning-Based Optimal Cooperating Node Selection for IoUT. Employs ML algorithms for selecting cooperating nodes based on delay, energy, and collision rates. & Uses DDPG-SEC algorithm for improved EE, reduced latency, and enhanced packet delivery, showing significant advancements over traditional methods. & EC, PDR, throughput, NL, successful transmission probability, and E2E delay. \\ 

\hline
2024 \cite{tarif2024_fuzzy} & Enhancing Energy Efficiency of Underwater Sensor Network Routing to Achieve Reliability using a Fuzzy Logic-based Approach. Implements a clustering-based routing method utilizing fuzzy logic to optimize energy consumption and reliability by considering factors like residual energy, distance, depth, and number of neighbors for node selection. &
Efficiently reduces energy consumption and improves network reliability. Balances traffic load and extends the network lifespan through dynamic clustering and fuzzy logic. & Residual energy, Distance to sink, Depth, Number of neighbors, Packet generation rate, Network topology, Communication range.\\ 
\hline

2024 \cite{tarif2024proposing} & To improve routing efficiency in underwater IoT networks by dynamically weighing routing parameters and enabling optimal distributed decision-making among network components. & The method enhances network lifetime, increases packet delivery rates, and reduces end-to-end latency through a multi-criteria decision-making system and multi-path routing. &   hops to root, node depth, ARSSI rate, energy, PDR link, ETX rate, delay, and node sector.\\

\hline
2024 \cite{Saemi2024_77} & Energy-efficient routing protocol using a hybrid metaheuristic algorithm (GSLS) for UWSNs. Combines Global Search Algorithm (GSA) and Local Search Algorithm (LSA) for optimal routing paths. & Efficiently reduces energy consumption and routing discovery time by leveraging a parallel search mechanism, significantly improving UWSN performance. & EC, PDR, NL, algorithm speed. \\ 
\hline
\end{tabular}
}
\end{table*}

\newpage
\clearpage

\section{NETWORK SETTINGS} \label{sec3}
The RPL protocol is designed and developed for the WSN and LLN networks in the IoT infrastructure. Changes to the protocol structure are necessary to implement it in an underwater environment. According to the structural differences between the wireless sensor network and the underwater audio sensor network, these differences will be applied in different network layers. Some of the most important adjustments are presented in sections 3.1 to 3.6. These changes are modelled in RPL code in the NS2 environment and the Aquasim package.

\subsection{Speed of sound in water}
The measure of the speed of sound frequency in the sea environment ($\zeta$) is related to three primary factors called water temperature ($T$), salinity ($\psi$), and depth ($d$). According to Mackenzie's formula \cite{Mackenzie1981_78}, a relatively accurate estimate of the speed of sound frequency underwater can be obtained by Equation~\ref{eq:Speed-of-Sound}, \cite{Leroy2008_79}.

\begin{equation} \label{eq:Speed-of-Sound}
\begin{split}
\zeta = & (1449 + 4.591T) - (5.304 \times 10^{-2}T^2) \\
& + (2.374 \times 10^{-4}T^3) + (1.34(\psi - 35)) \\
& + (1.63 \times 10^{-2}d) + (1.675 \times 10^{-7}d^2) \\
& + (1.025 \times 10^{-2}T(\psi - 35)) - (7.139 \times 10^{-3}Td^3)
\end{split}
\end{equation}

\subsection{Underwater frequency link quality criterion}
Factors affecting the quality of the underwater connection include noise due to water turbulence $N_t$ and noise due to the movement of vessels with a coefficient $\epsilon$ denoted by $N_s$. Also, $N_w$ is the noise caused by the waves generated by the wind at a speed of m/s, and finally, $N_{th}$ is the ambient thermal noise. The spectral density power of a frequency in an underwater environment is calculated through Equations~\ref{eq:N_total} to~\ref{eq:N_thermal} \cite{Shahapur2021_80}:

\begin{equation} \label{eq:N_total}
N(f) = N_t(f) + N_s(f) + N_w(f) + N_{th}(f)
\end{equation}

\begin{equation} \label{eq:N_turbulence}
10 \log N_t(f) = 17 - 30 \log f
\end{equation}

\begin{equation} \label{eq:N_ships}
10 \log N_s(f) = 40 + 20 (e - 0.5) + 26 \log f - 60 \log (f + 0.03)
\end{equation}

\begin{equation} \label{eq:N_waves}
10 \log N_w(f) = 50 + 7.5 \sqrt{w} + 20 \log f - 40 \log (f + 0.4)
\end{equation}

\begin{equation} \label{eq:N_thermal}
10 \log N_{th} (f) = -15 + 20 \log f
\end{equation}

In these equations, $\epsilon$ is equal to the coefficient of the noise of the shipping factor, which takes values between 0 and 1. Also, $w$ equals the wind speed, which varies from 0 to 10 meters per second. It is observed that the noise rate of the carrier frequency $N(f)$, given by Equation~\ref{eq:N_total}, increases with increasing the amount of $\epsilon$ and $w$ in the environment. The signal-to-noise ratio, known as SNR, is calculated by Equation~\ref{eq:SNR} \cite{Ryecroft2019_81}:

\begin{equation} \label{eq:SNR}
SNR = \frac{P}{N(f)}
\end{equation}

In this relation, $P$ is equal to the power of the transmitted signal in a narrow frequency band. According to Shannon's theory, the communication channel's capacity $C$, is given by Equation~\ref{eq:channel_capacity}.

\begin{equation} \label{eq:channel_capacity}
C(d,f) = B \log_2(1 + SNR(d,f))
\end{equation}

According to network logic, as the noise rate in the media environment increases, the available capacity of the channel decreases. This is represented by $N(f)$ in Equation~\ref{eq:SNR}, based on the link stability rate. Thus, $(d,f)$ can be used to gauge both link quality and as a benchmark for detecting acceptable levels of noise communication in underwater networks.

\subsection{Delay Time model}
Underwater sensor nodes use sound waves to transmit information, and the speed of underwater sound is about 1500 meters per second. This value is several times less than radio signals in the out-of-water environment \cite{Liu2021_82}. Therefore, the signal propagation delay in the underwater environment is significant. Total underwater delays include processing time $DT_{proc}$, queue delay $DT_{queue}$, propagation delay $DT_{prop}$, and transmission delay $DT_{trans}$.

\begin{equation} \label{eq:prop_delay}
DT_{prop} = \sum_{i=1}^{HC} \frac{d_i}{v_p}
\end{equation}

\begin{equation} \label{eq:trans_delay}
DT_{trans} = HC \left( \frac{L}{R_{bit}} \right)
\end{equation}

In relations~\ref{eq:prop_delay} and~\ref{eq:trans_delay}, the parameter $d_i$ equals distance, $v_p$ equals signal propagation speed, and $HC$ is the hop-count. The processing and queue latency is much less than the propagation delay, and often, in calculations, they deal with only the two propagation and transmission delays \cite{Homaei2019_83}. Finally, the total latency of a packet from origin to destination, passing through $n$ hops, is calculated by Equation~\ref{eq:total_delay}.

\begin{equation} \label{eq:total_delay}
DT_{all} = \sum_{i=1}^{N} (DT_{proc} + DT_{queue} + DT_{prop} + DT_{trans})
\end{equation}

\subsection{Calculating node depth}
Each underwater sensor node is equipped with a depth gauge unit to obtain the amount of pressure applied to the node, as indicated by \(P\). The value of \(P\) is calculated from the equation \(P = \rho g\) where \(\rho\) and \(g\) are two computational constant values. The depth difference between nodes A and B, denoted by \(\Delta d\), is calculated by Equation~\ref{eq:depth_difference}. Here, \(d_A\) represents the depth of the sender node, and \(d_B\) is the depth of the sender node's neighbor.

\begin{equation} \label{eq:depth_difference}
\Delta d = d_A - d_B = \frac{P_A - P_B}{\rho g}
\end{equation}

\subsection{Frequency attenuation or absorption model}

The \(\alpha\) parameter, measured in dB/km, is used to quantify the rate of frequency absorption. Typically, the power of sound frequency underwater diminishes by about 21\% per kilometer. The depth of water plays a significant role in determining the attenuation rate \cite{Ainslie1998_84}.

\begin{equation} \label{eq:attenuation_at_depth}
\alpha_d = \alpha_0 (1 - 1.93 \times 10^{-5} d)
\end{equation}

Equation~\ref{eq:attenuation_at_depth} describes the degree of attenuation at a certain depth \(d\), where \(\alpha_0\) is the attenuation at the surface level. As the equation indicates, the absorption rate decreases in deeper water.

\begin{multline} \label{eq:frequency_absorption}
\alpha = 0.106 \frac{f_1 \times f^2}{f_1^2+f^2}  e^\frac {\text{pH}-8}{\text{e}^{0.56}} + 0.52 \left(1 + \frac{T}{43}\right) \\
\times (\frac{\Psi}{25})\frac{f_2 \times f^2}{f_2^2 + f^2}\text{e}^\frac {-d}{6} + \text{4.9} \left(10^{-4}f^2 e^{-\left(\frac{T}{27} + \frac{d}{17}\right)}
\right)
\end{multline}

Where \(\text{pH}\) represents the water's acidity, \(\Psi\) its salinity, \(T\) the temperature in Celsius, and \(d\) the depth in kilometers. The frequencies \(f_1\) and \(f_2\) relate to the water's salinity and temperature, respectively, as defined in Equations~\ref{eq:frequency_f1} and~\ref{eq:frequency_f2}.

\begin{equation} \label{eq:frequency_f1}
f_1 = 0.78 \sqrt{\frac{\Psi}{35}} \left(e^{\frac{T}{26}}\right)
\end{equation}

\begin{equation} \label{eq:frequency_f2}
f_2 = 42 \left(e^{\frac{T}{17}}\right)
\end{equation}

Equation~\ref{eq:frequency_absorption}, for simplicity in calculations, assumes a temperature of \( 4 ^{\circ}C \), a depth of 1000m, a salinity of 30, and a \( \text{pH} \) level of 8.

\subsection{Structure of control packets}

The RPL network uses four control packets to form, repair, and maintain the network graph. Some messages are sent periodically, while others are dispatched as needed in response to network errors or instability. These messages have been redefined and adapted for the underwater sensor network within the proposed RPLUW protocol.

\begin{itemize}
    \item The DIO packet is broadcasted by parents to the children to promote DODAG.
    \item The DAO packet is unicast from membership by the child to the parent.
    \item The DAO-Ack packet is also unicast from the parent to the child to confirm the membership of the child node in the parent list.
    \item If the node is orphaned or located in an area where the network instability has reached the threshold, the DIS packet will be broadcasted by the parentless node in the network.
\end{itemize}

\begin{figure}[h]
    \centering
    \includegraphics[width=0.4\textwidth]{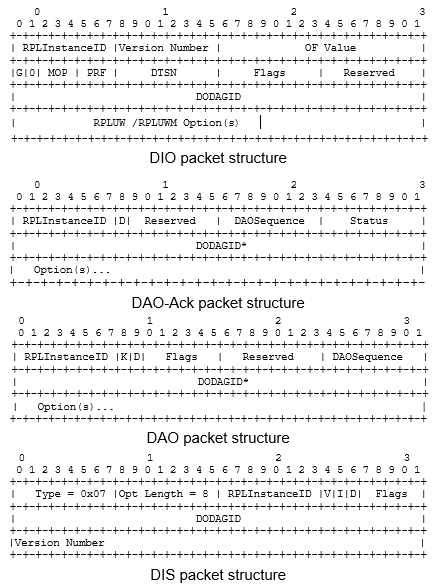}
    \caption{The structure of control packets in RPLUW and RPLUWM}
    \label{fig:control_packets_structure}
\end{figure}

\section{THE PROPOSED RPLUW METHOD} \label{sec4}
\subsection{The system model}
\begin{figure}[h]
    \centering
    \includegraphics[width=1.0\linewidth]{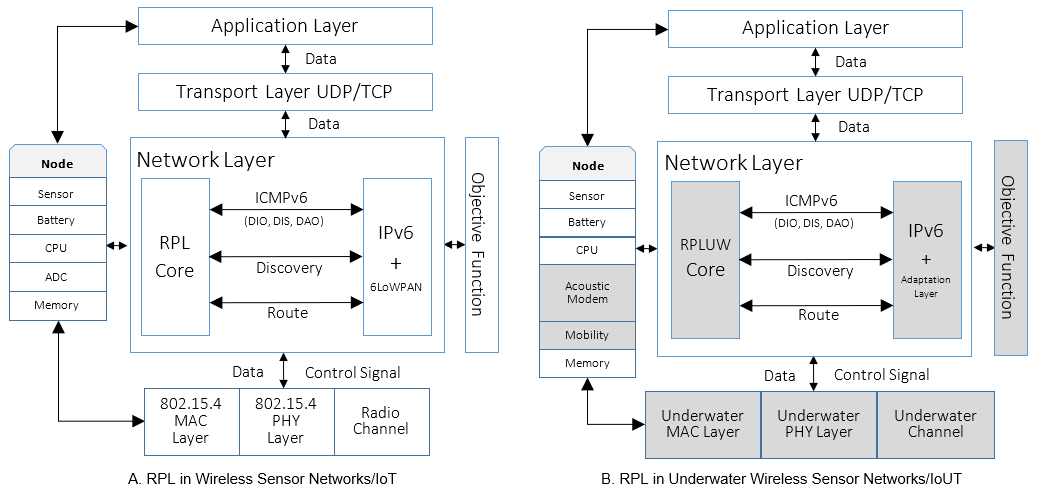}
    \caption{Adaptation of the RPL protocol in an underwater environment}
    \label{fig:rpl_protocol_structure}
\end{figure}

RPL is an IPv6 distance vector routing protocol that operates on the physical layers and IEEE 802.15.4 data link and is suitable for sensor networks with low power sources and minimal bandwidth. With the advancement of technology, more appropriate facilities and communication platforms have been provided for these networks than in the previous decade. In this paper, we present a new development of RPL called RPLUW, which is the motionless version, and RPLUWM, the version that supports mobility in underwater sensor networks. To provide these methods, significant structural changes were necessary to make the RPL compatible with the underwater environment. The significant changes to the proposed method with the standard version of RPL are shown in gray in Figure \ref{fig:rpl_protocol_structure}.

The following are some definitions and details of the proposed protocol.

\begin{itemize}
    \item The rank \( R(u, j) \) is a measure that defines the logical distance of node \( u \) as a subset of \( N (u \in N) \) from the root of the network graph \( J \), by the objective function (OF) of the RPL protocol. The \( R \) rate will often increase as it moves away from the graph's root. In the underwater network, this criterion can be defined according to the application, such as the amount of node depth from the water surface or as a combination of step and node depth.
    \item The Preferred Parent DODAG (DPP): Suppose \( u \) is a node of \( G \), \( N(u) \) its one-hop neighbors, and \( DPP(u,j) \) a finite subset of \( N(u) \). For each node \( v \in N(u) \), we have \( v \in DPP(u, j) \), if it has the lowest rank up to the root \( R(u, j) \) specified DODAG \( j \in B \). In the improved version of RPL, we will have several preferred parents for each node, and the best parent will be chosen by node \( u \) when the node sends the packet.
    \item DODAG (DRL) root list: Each node \( v \in N(u) \) must broadcast DIO packets. The root location of DODAG should be included in these packets. Thus, \( DRL(u) \) is a list of DODAG root locations stored at each node \( u \) in the network.
\end{itemize}


\begin{figure}[h]
    \centering
    \includegraphics[width=0.5\linewidth]{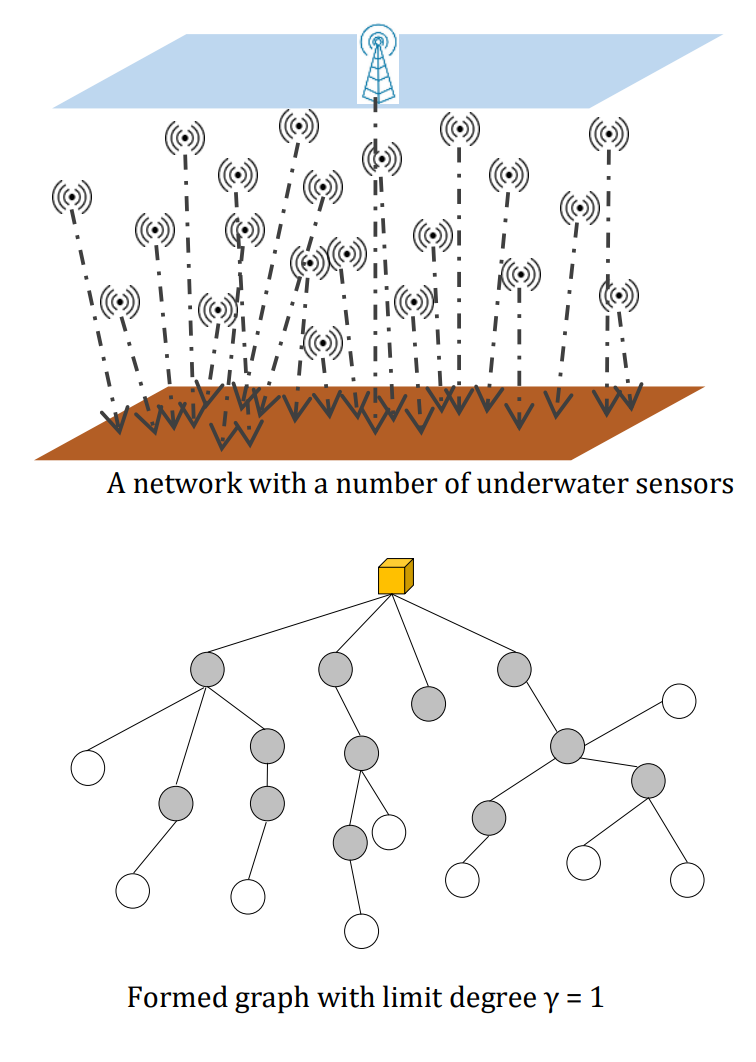}
\caption{RPL network graph in underwater conditions, unrestricted and with \(\gamma = 1\) child node limit.}
\label{fig:rpl_network_graph}
\end{figure}

The proposed Underwater-RPL method forms and supports at least one DODAG per well node. This protocol calculates the upward and downward paths independently and according to a process to benefit from them if placed in the network objective function. In the RPL method, the base of each node is bound to have a preferred parent in its list, which directs the generated data and data received from its offspring to the root by the preferred parent. In this method, the network is modelled as a \( G (V, E) \) directionless link graph. A constant \( k \) is defined where \( \gamma = 1 \) is the preferred parent number of each node in the network. This \( \gamma = 1 \) method in the basic RPL protocol suffers from some injustice and imbalance in traffic transfer because each parent node can have \( n \leq N-1 \) members in the worst case. This causes imbalance, queue inefficiency, and congestion in the network. Therefore, according to the new RPL models, we use a multi-path mechanism in the proposed method. That is, for each node \( u \in G (V, E) \), the value of \( k \geq 1 \). To better understand this multi-path mechanism, suppose that, according to Figure\ref{fig:rpl_network_graph}, several nodes are randomly distributed in the underwater environment. The sink node is located on the water surface and is connected to the water floor by anchors and chains like the network nodes.

\begin{figure}
    \centering
    \includegraphics[width=0.5\linewidth]{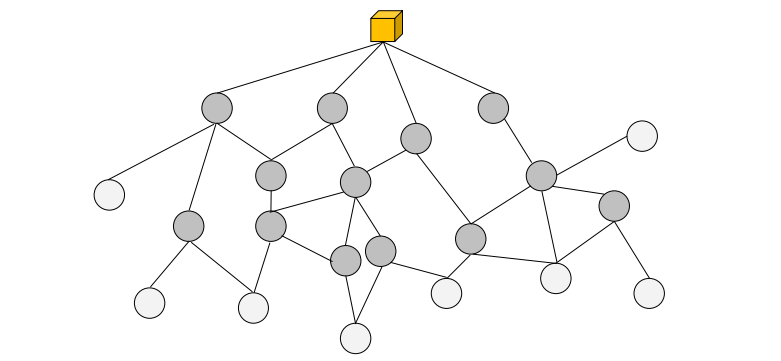}
    \caption{Demonstration of a network graph without parental degree constraints}
    \label{fig:rpl-graph-without-limit-degree}
\end{figure}

Figure \ref{fig:rpl-graph-without-limit-degree} shows that we will face the following graph by removing the limit value of the number of parents for the child node. The number of available parents for each child node is \( k \geq 1 \), which increases the number of network paths and reduces network failure.

\subsection{Detection and management of displacement in underwater RPL}

Displacement is one of the main reasons for graph incompatibility and instability in RPL. Each node in the network is moved or removed from its neighbor's list, and its neighbor list changes periodically. In general, there are two main methods for detecting and managing displacement:

\begin{itemize}
    \item \textbf{Trickle timer algorithm:} A dynamic schedule based on network stability/instability rate prevents repetitive sending of control packets and keeps the network graph's repair mechanism dynamic. The trickle timer of the t-period is limited by the interval \([I_{\text{min}}, I_{\text{max}}]\), where \(I_{\text{min}}\) is the minimum interval defined in milliseconds by the base value of 2 (e.g., \(2^{12} = 4096\) ms) and \(I_{\text{max}} = I_{\text{min}} \times 2^{I_{\text{doubling}}}\) is used to limit the number of times \(I_{\text{min}}\) can be doubled. Assuming \(I_{\text{doubling}} = 4\), the maximum interval is calculated as \(I_{\text{max}} = 4096 \times 2^{4} = 65536\)ms. The trickle timer algorithm can update the topology in a short time. When the incompatibilities of 13 nodes are reached, i.e., \(IC \geq k\), it sets the value of \(t\) equal to \(I_{\text{min}}\) and updates the tree. If the network is stable, i.e., \(IC < k\), the value \(I\) double until the constant value \(I_{\text{max}}\) is reached. If the network nodes are moved, the number of DIS packets in the network increases, and this causes the \(IC\) to converge to the value of \(k\), and the scheduler returns to \(I_{\text{min}}\). As the timing of DIO packet periods in the network decreases, the overhead network will increase, and the network will be involved in resolving incompatibilities.
\end{itemize}

\begin{itemize}
    \item IPv6 Neighbour discovery method: RPL can use neighbor discovery to detect environmental changes with an optimal version of ND. ND makes it possible to detect neighbor inaccessibility and discover new neighbors, which is supported by four ICMPv6 control packets:
    \begin{itemize}
        \item Neighbour Solicitation (NS): Checks the node's availability by checking the neighbor's MAC address.
        \item Neighbour Advertisement (NA): Responds to NS packets, is also sent intermittently to announce link changes.
        \item Router Solicitation (RS): The host node (mobile node in the proposed model) solicits information from the router.
        \item Router Advertisement (RA): The router periodically sends its presence packet and graph and link parameter information to respond to the RS packet.
    \end{itemize}
\end{itemize}

\subsection{The RPLUWM method timers}

We used several timers to increase underwater network communications' stability, availability, and reliability. These timers are designed to control the process of reducing link instability and node inaccessibility to the network's parent.

\begin{itemize}
    \item \textbf{Linkage timer (\(L_t\)):} To increase the response of network nodes, they must periodically monitor the activity of the communication channel. An \(L_t\) timer is installed for this process, the sequence value of which is determined by a trickle timer (\(I_{\text{max}}\)). During this process, the moving node listens to the channel and monitors the input packets from the parent. After the \(L_t\) timer expires, the discovery phase begins if the parent exchanges reach zero. Also, the link timer is reset upon receipt of any packet from the parent (such as trickle DIO, unicast DIO, or data packet).
    \item \textbf{Mobility timer (\(M_t\)):} After receiving the unicast DIO packet, the quality of the Average received signal strength indicator (ARSSI) is evaluated to evaluate the reliability of the link. As the ARSSI rate decreases, the moving node begins to explore to find a new parent. The node data generation rate sets this timer during network setup.
    \item \textbf{Response timer (\(R_t\)):} If the parent detects disconnection and receives a DIS packet from the neighbors, it must send a DIO packet at certain times. Selecting the wrong response moment may cause data packets to collide, activating the exploration phase. The parent node monitors the other children's packets, estimates the packets' sequence and response time, and sends a DIO packet outside.
\end{itemize}

\begin{figure}
    \centering
    \includegraphics[width=0.8\linewidth]{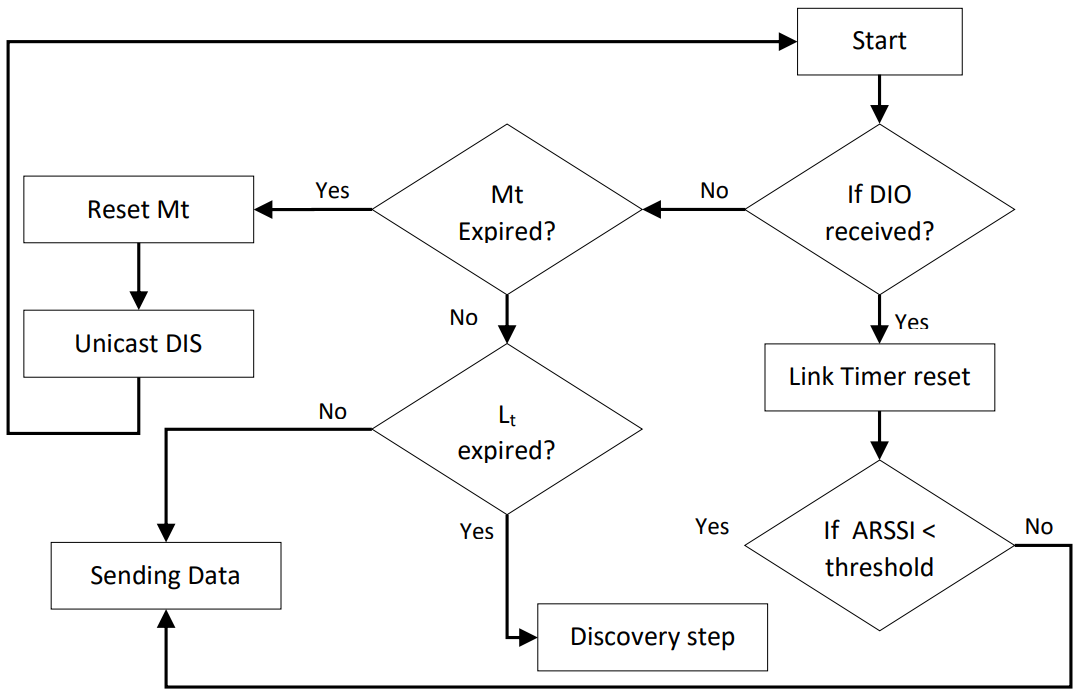}
    \caption{RPLUW network scheduling steps}
    \label{fig:RPLUW_network_scheduling}
\end{figure}

\subsection{RPLUWM network graph construction method}

The network graph formation steps in the RPLUWM protocol are created using the proposed objective function. In the basic RPL protocol, the hop or rank of the nodes is usually the criterion for graph formation. However, in the proposed method, we have used a combination of three parameters: hops to root, node depth from the water level, and ARSSI rate of the interface link to form the initial graph (Algorithm \ref{Algorithm1}). 

\begin{algorithm}
\caption{RPLUWM Protocol Network Graph Formation}
\begin{algorithmic}[1]
\State \textbf{Input:} Underwater network environment with nodes, including a sink node
\State \textbf{Output:} Network graph formed based on an objective function

\State The sink node broadcasts a DIO packet to all nodes in the network
\For{each node receiving a DIO packet}
    \State Send a DAO packet back to the sink node with the node's depth and link ARSSI rate
\EndFor
\State Sink node sends DAO-Ack packets to first-level nodes due to unlimited one-hop children
\For{each parent node}
    \State Update hop value, current depth, and ARSSI rate in the DIO packet
    \State Resend the updated DIO packet to other nodes
\EndFor
\For{each node receiving updated DIO packets}
    \State Compile and update a list of potential parents based on received values
\EndFor
\For{each child node}
    \State Select one or more parents from the list and send a DAO request with depth and ARSSI
\EndFor
\For{each parent node receiving DAO packets}
    \If{link quality level to the child node is above the threshold}
        \State Send a DAO-Ack packet to the child node
    \EndIf
\EndFor
\State Child nodes receiving DAO-Ack packets proceed to form a graph, continue to the next hop
\State Repeat the process until the entire network graph is formed
\end{algorithmic}\label{Algorithm1}
\end{algorithm}

\subsection{Routing in the RPLUWM network}

Routing is the central pillar of data transfer in the Internet of Things. This means that without a logical mechanism that is aware of the resources and capacity of the network, data transfer is associated with a high resource consumption overhead or, in some applications, seems impossible. In the Internet of Things, multi-path mechanisms are used to improve network reliability and minimize bottlenecks or hotspots. In this section, we propose a quality service-aware approach in the underwater mobile sensor network that can achieve acceptable efficiency with minimal consumption of network resources. To this end, an optimal multi-criteria decision system \cite{Homaei2019_83} proposed in the previous work for the RPL network in the offshore environment with changes and adaptation to the underwater network has been proposed for routing. This design can combine all the effective parameters according to the network's needs and determine each parent's final value and priority for the nodes. After forming the network graph in the first step, each node has a list in its memory, the values of which will be updated periodically. This list is assigned to the status of available parents for the node whose maximum number of parents in this list is limited to \textit{\textbf{$\bar{k}$ = 4}}; Because, according to tests, more than a fixed number of parents per node increases congestion, hidden terminals, and requires more complex calculations. Table \ref{tab:example-parameters-of-madm} shows this list of parents.

\begin{table}[h!]
    \centering
    \scriptsize
    \caption{An example of a list of parents of each node with their quantitative values}
    \label{tab:example-parameters-of-madm}
    \begin{tabular}{|p{70pt}|p{25pt}|p{25pt}|p{25pt}|p{25pt}|}
        \hline
        Parents / Parameters & $P_i$ & $P_{i+1}$ & $P_{i+2}$ & $P_{(i \leq \bar{k})}$ \\ 
        \hline       
        Hop-Count$(n)$ & 3 & 3 & 3 & 3 \\
        Remaining Energy$(j)$ & 167.5 & 183.2 & 179 & 138.8 \\
        ARSSI & $\bar{R}_i$ & $\bar{R}_{i+1}$ & $\bar{R}_{i+2}$ & $\bar{R}_{i \leq \bar{k}}$ \\
        Delay Time(ms) & $DT_i$ & $DT_{i+1}$ & $DT_{i+2}$ & $DT_{i \leq \bar{k}}$ \\
        ETX & $\epsilon_i$ & $\epsilon_{i+1}$ & $\epsilon_{i+2}$ & $\epsilon_{i \leq \bar{k}}$ \\
        Link's PDR (\%) & 0.78 & 0.85 & 0.76 & 0.88 \\ 
        Depth(m) & 129.8 & 141.2 & 155.4 & 117.4 \\ 
        \hline   
    \end{tabular}
\end{table}

\begin{algorithm}
\setcounter{ALG@line}{0}
\caption{Multi-Criteria Decision Making AHP Matrix}
\begin{algorithmic}[1]
\State \textbf{Input:} Pairwise comparison matrix $A$ of size $n \times n$
\State \textbf{Output:} Priority vector $\omega$ of size $n$

\State Initialize sum vector $S$ of size $n$ to zeros
\State Initialize priority vector $\omega$ of size $n$ to zeros

\For{$j = 1$ to $n$} \Comment{Calculate column sums}
    \State $columnSum[j] \gets \sum_{i=1}^{n} A[i][j]$
\EndFor

\For{$i = 1$ to $n$} \Comment{Normalize the matrix}
    \For{$j = 1$ to $n$}
        \State $A[i][j] \gets A[i][j] / columnSum[j]$
    \EndFor
\EndFor

\For{$i = 1$ to $n$} \Comment{Calculate priority vector}
    \State $S[i] \gets \sum_{j=1}^{n} A[i][j]$
    \State $\omega[i] \gets S[i] / n$
\EndFor

\State \Return $\omega$
\end{algorithmic} \label{Algorithm2}
\end{algorithm}

In the section on prioritizing parent nodes, a multi-criteria decision-making system was used for children to make optimal selections \ref{Algorithm2}. In this approach, each parent node is weighted based on the parameters of the hop, energy, ARSSI rate, latency, ETX, link delivery rate, and depth in the decision system and will obtain its final value in combination. In the next matrices, the steps for calculating the decision system are given.
For example:

\[
A = \begin{bmatrix}
1      & a_{12} & a_{13} & \cdots & a_{17} \\
1/a_{12} & 1      & a_{23} & \cdots & a_{27} \\
\vdots & \vdots & \vdots & \ddots & \vdots \\
1/a_{17} & 1/a_{27} & 1/a_{37} & \cdots & 1
\end{bmatrix}
\]

\[
\begin{array}{c}
\Big\Downarrow \\
\end{array}
\]

\[
\left[
\begin{array}{c|cccc|c}
     & x_1    & \cdots &    \cdots  & x_n & S \\
\hline
x_1  & 1      & \cdots & \cdots &  \frac{\omega_1}{\omega_n} & \sum_{i=1}^{n}\omega_{1,i} \\
\vdots & \vdots & \ddots &  & \vdots  & \vdots  \\
\vdots & \vdots &  & \ddots & \vdots & \vdots \\
x_n  & \frac{\omega_n}{\omega_1} & \cdots &\cdots & 1 & \sum_{j=1}^{n}\omega_{i,j} \\
\end{array}
\right]
\]
\[
\begin{array}{c}
\Big\Downarrow \\
\end{array}
\]
\[
\left[
\begin{array}{c|cccc||c}
     & x_1    & \cdots &    \cdots  & x_n & S \\
\hline
x_1  & 1 & \cdots & \cdots & \frac{\omega_1}{\omega_n} & \sum_{i=1}^{n}\omega_{1,i} \\
\vdots & \vdots & \ddots &  & \vdots  & \vdots  \\
\vdots & \vdots &  & \ddots & \vdots & \vdots \\
x_n  & \frac{\omega_n}{\omega_1} & \cdots &\cdots & 1 & \sum_{i=1}^{n}\omega_{n,i} \\
\hline
S    & \sum_{j=1}^{n}\omega_{j,1} & \cdots & \cdots & \sum_{j=1}^{n}\omega_{j,n} & \sum_{i,j=1}^{n}\omega_{i,j}
\end{array}
\right]
\]   
\[
\begin{array}{c}
\Big\Downarrow \\
 \end{array}
\]
\[
\left[
\begin{array}{c|ccccc}
 & x_1    & \cdots &    \cdots  & x_n & S \\
\hline
x_1  & \frac{1}{1}      & \frac{\omega_1}{\omega_2} & \cdots &  \frac{\omega_1}{\omega_n} & \frac{\sum_{j=1}^{n}\omega_{1,j}}{n} \\
\vdots & \frac{\omega_2}{\omega_1} & \frac{1}{1} & \cdots & \frac{\omega_2}{\omega_n} & \frac{\sum_{j=1}^{n}\omega_{2,j}}{n} \\
\vdots & \vdots & \vdots & \ddots & \vdots & \vdots \\
x_n  & \sum_{i=1}^{n}\frac{\omega_{i,1}}{\omega_{something}} & \cdots &\cdots & \frac{1}{1} & \frac{\sum_{j=1}^{n}\omega_{n,j}}{n} \\
\end{array}
\right]
\]
\[
\begin{array}{c}
\Big\Downarrow \\
 \end{array}
\]

\noindent 
\[
\begin{bmatrix}
\frac{\sum_{j=1}^{n}\omega_{1,j}}{n} \\
\frac{\sum_{j=1}^{n}\omega_{2,j}}{n} \\
\vdots \\
\frac{\sum_{j=1}^{n}\omega_{n,j}}{n}
\end{bmatrix}
\Rightarrow
\begin{bmatrix}
\omega_1 \\
\cdot \\
\cdot \\
\cdot \\
\omega_n
\end{bmatrix}
\Rightarrow
\begin{bmatrix}
0.48 \\
0.24 \\
0.16 \\
0.08 \\
0.04
\end{bmatrix}
\]

Finally, the values for each parent node in the children list will be updated, and if the nodes move, the ARSSI value will be updated separately at a higher sampling rate. Equations \ref{eq:Nodevalue} and \ref{eq:MADMSelection} decide how to calculate the system.

The value of each node can be calculated as follows:
\begin{equation} \label{eq:Nodevalue}
\text{NodeValue}(k = 1 \ldots n) = \sum_{k=1}^{n} (\text{Param}_k \times \omega_k)
\end{equation}

The MADM selection criterion is given by:
\begin{equation} \label{eq:MADMSelection}
\text{MADM selection} = \max (\text{NodeValue}_k)
\end{equation}

Section 5 simulates and evaluates the proposed method against recent methods.

\section{SYSTEM MODEL AND SIMULATION} \label{sec5}

To evaluate and compare the proposed method with the recent methods of EEGNBR \cite{Sathish2022_52}, Co-DRAR \cite{Ullah2023_65}, OR-RTP \cite{Simon2023_70}, and UA-RPL \cite{Saravanan2022_63}, NS simulation version 2.31 and Aquasim package version 2 were used. The beam widths of each underwater sensor node varied between 0 and 360 degrees. The radio range of the sensor node was 150 meters and the radio range of the well was 200 meters. Network nodes were randomly distributed in the underwater environment. This section evaluates all the details and simulation conditions in Table \ref{tab:network-conditions}.

\begin{table}[ht]
\centering
\caption{Network simulation conditions}
\label{tab:network-conditions}
\setlength{\tabcolsep}{3pt}
\begin{tabular}{|p{123pt}|p{110pt}|}
\hline
\textbf{Parameters} & \textbf{Value} \\
\hline
Network topology & Random position \\
Deployment area & 1000 x 500 m\textsuperscript{3} \\
Initial node energy & 50 J \\
Initial sink energy & 50kJ \\
Number of nodes & 50, 100, 200 \\
Nodes mobility & 1 m/s–5 m/s \\
Mobility model & Random mobility \\
Percentage of Mobile Nodes & 40\% \\
Cost of long transmission & 1.3 W \\
Cost of short transmission & 0.8 W \\
Cost of reception & 0.7 W \\
Idle power & 0.008 W \\
Data aggregation power & 0.22 W \\
Communication range of ASN & 150 m \\
Acoustic transmission range(sink) & 200 m \\
Spreading values & 1.3 \\
Frequency & 30.5 kHz \\
Channel & Underwater channel \\
Maximum Bandwidth & 30 kbps \\
DIO packet size & 50 bytes \\
DAO packet size & 4 bytes \\
DAO-ACK packet size & 4 bytes \\
DIS packet size & 4 byte \\
Packet generation rate & $\lambda = 0.1\sim0.2$ pkt/s \\
Memory size & 12 MB \\
Sink position & Surface (500 x 500 x 0) \\
Antenna & Omni-directional \\
Simulation time & 600 \\
Iterations & 10 \\
Number of Channels & 11 (30.511, 30.518, ... 30.581) kHz \\
\hline
\multicolumn{2}{p{243pt}}{*Bellhop is used to calculate the path loss between each node in a given location.} \\

\end{tabular}
\end{table}

\subsection{Testing the effect of environment on network energy consumption rate}
\begin{figure}[H]
    \centering
    \includegraphics[width=0.6\linewidth]{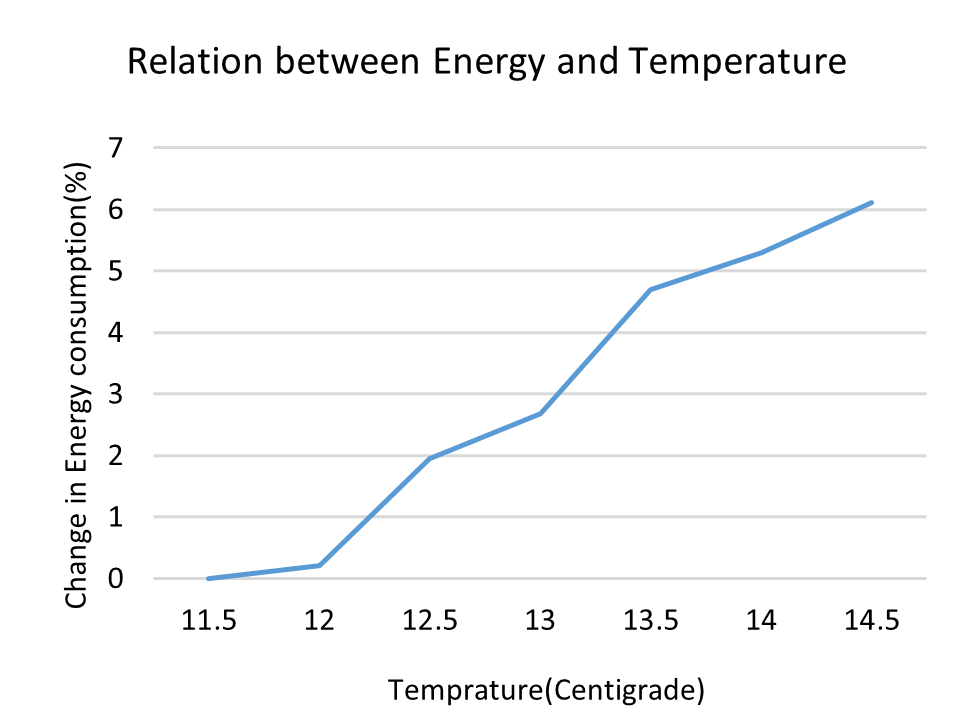} 
    \caption{The rate of increase in node energy consumption under varying temperatures.}
    \label{fig:temperature}
\end{figure}

\begin{figure}[H]
    \centering
    \includegraphics[width=0.6\linewidth]{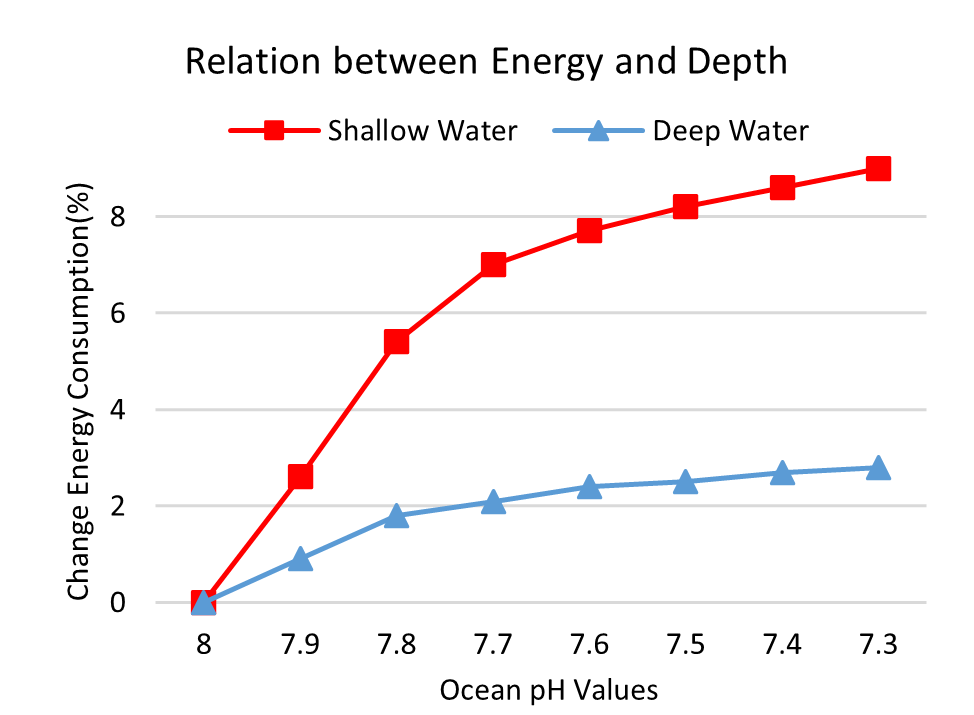} 
    \caption{The changes in energy consumption increase for varying degrees of salinity in shallow and deep water environments.}
    \label{fig:salinity}
\end{figure}

To evaluate the energy consumption rate, it is not enough to pay attention to the hardware and software energy model of the underwater sensor nodes. Instead, attention should be paid to the water temperature, ambient salinity rate, and node depth for network exchanges. In Figure~\ref{fig:temperature}, we model the rate of increase in node energy consumption under temperatures ranging from \(11.5\,^{\circ}\mathrm{C}\) to \(14.5\,^{\circ}\mathrm{C}\). Due to the lack of temperature changes in deep water environments, energy consumption will not change significantly. Also, in Figure~\ref{fig:salinity}, the changes in energy consumption increase are modeled for two shallow and deep water environments with varying degrees of salinity, and the obtained results are plotted.

\subsection{Network lifetime test}
The viability of underwater networks hinges critically on the network's lifespan, as the exhaustion of node energy denotes the cessation of connectivity and, subsequently, the surveillance operation's failure. Standard benchmarks for a network's longevity include the times to failure of the first and median nodes, with imbalanced energy consumption accelerating these events. The RPLUW technique and its mobile adaptation, RPLUWM, have demonstrated enhanced outcomes by intelligently employing network graphs and prioritizing quality of service in routing. Moreover, the dynamic nature of the RPLUWM graph structure adeptly manages the inherent mobility of underwater networks, ensuring more rapid convergence than alternative methods. This swift convergence mitigates node disconnection, futile attempts, and energy depletion, prolonging network life and bolstering energy efficiency.

\begin{figure}[h!]
\centering
\includegraphics[width= 0.6\linewidth]{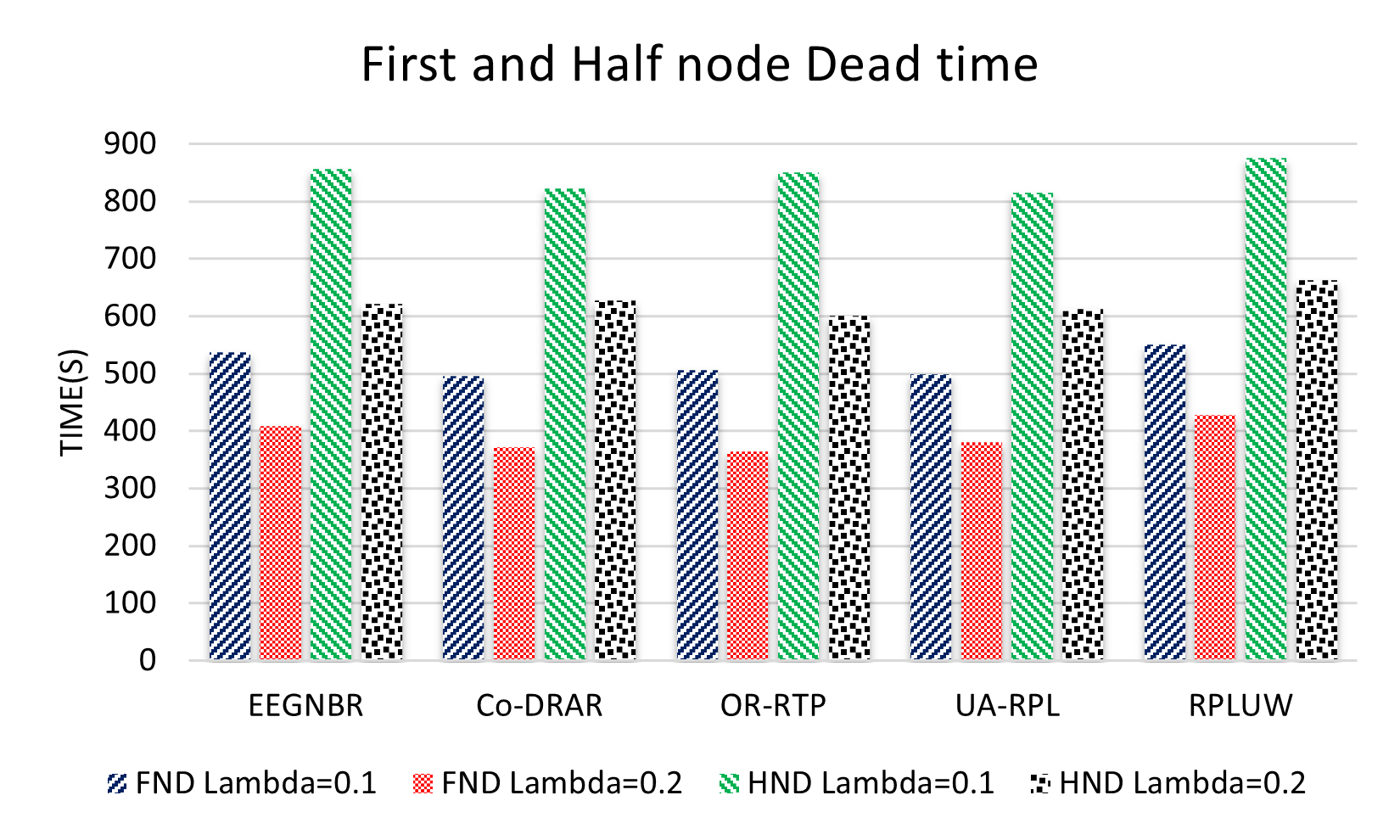}
\caption{Comparative analysis of the lifespan of the first and median nodes in underwater networks, showing the performance of the RPLUW method against other methods.}
\label{fig:node-death-time-static}
\end{figure}

Illustrated in Figure~\ref{fig:node-death-time-static}, the RPLUW method and its counterparts show distinct advantages in the longevity of the first and median network nodes. Specifically, at a traffic load of $\lambda = 0.1$ pkt/s, the RPLUW method extends the life of the first node by 2.36\%, 10.00\%, 8.00\%, and 9.45\% longer than EEGNBR, Co-DRAR, OR-RTP, and UA-RPL methods, respectively. When the traffic load is at $\lambda = 0.2$ pkt/s, the RPLUW's advantage is further pronounced, outperforming EEGNBR, Co-DRAR, OR-RTP, and UA-RPL by 4.67\%, 13.08\%, 14.95\%, and 11.21\% respectively. Regarding the median node lifespan, under a $\lambda = 0.1$ pkt/s traffic load, RPLUW outlasts EEGNBR, Co-DRAR, OR-RTP, and UA-RPL by margins of 2.28\%, 6.05\%, 2.97\%, and 6.96\% correspondingly. At a higher traffic load of $\lambda = 0.2$ pkt/s, RPLUW still maintains its lead, with lesser depletion rates than EEGNBR, Co-DRAR, OR-RTP, and UA-RPL by 6.33\%, 5.28\%, 9.50\%, and 7.69\% respectively.

In the dynamic realm of underwater networks, where mobility is inherent to many nodes, assessing the efficacy of routing protocols becomes critical. To this end, a refined iteration of the suggested method, RPLUWM, was rigorously tested against conventional methods such as EEGNBR, Co-DRAR, OR-RTP, and UA-RPL. The RPLUWM variant was designed to accommodate scenarios where around 40\% of the network nodes are mobile, a condition reflective of real-world applications and essential for a fair assessment of the method's robustness compared to others.

The core principle of RPLUWM, which integrates flexible scheduling and minimizes unproductive route explorations, significantly extended the operational lifetime of network nodes. When scrutinized at a traffic intensity of $\lambda = 0.1$ pkt/s, the RPLUWM approach surpassed its contemporaries, showing remarkable improvements in the longevity of the first node—EEGNBR by 9.51\%, Co-DRAR by 14.40\%, OR-RTP by 22.11\%, and UA-RPL by 14.14\%. For the median node, the increments in lifespan were equally noteworthy—EEGNBR by 11.03\%, Co-DRAR by 8.58\%, OR-RTP by 23.12\%, and UA-RPL by 10.72\%.

\begin{figure}[h!]
\centering
\includegraphics[width=0.6\linewidth]{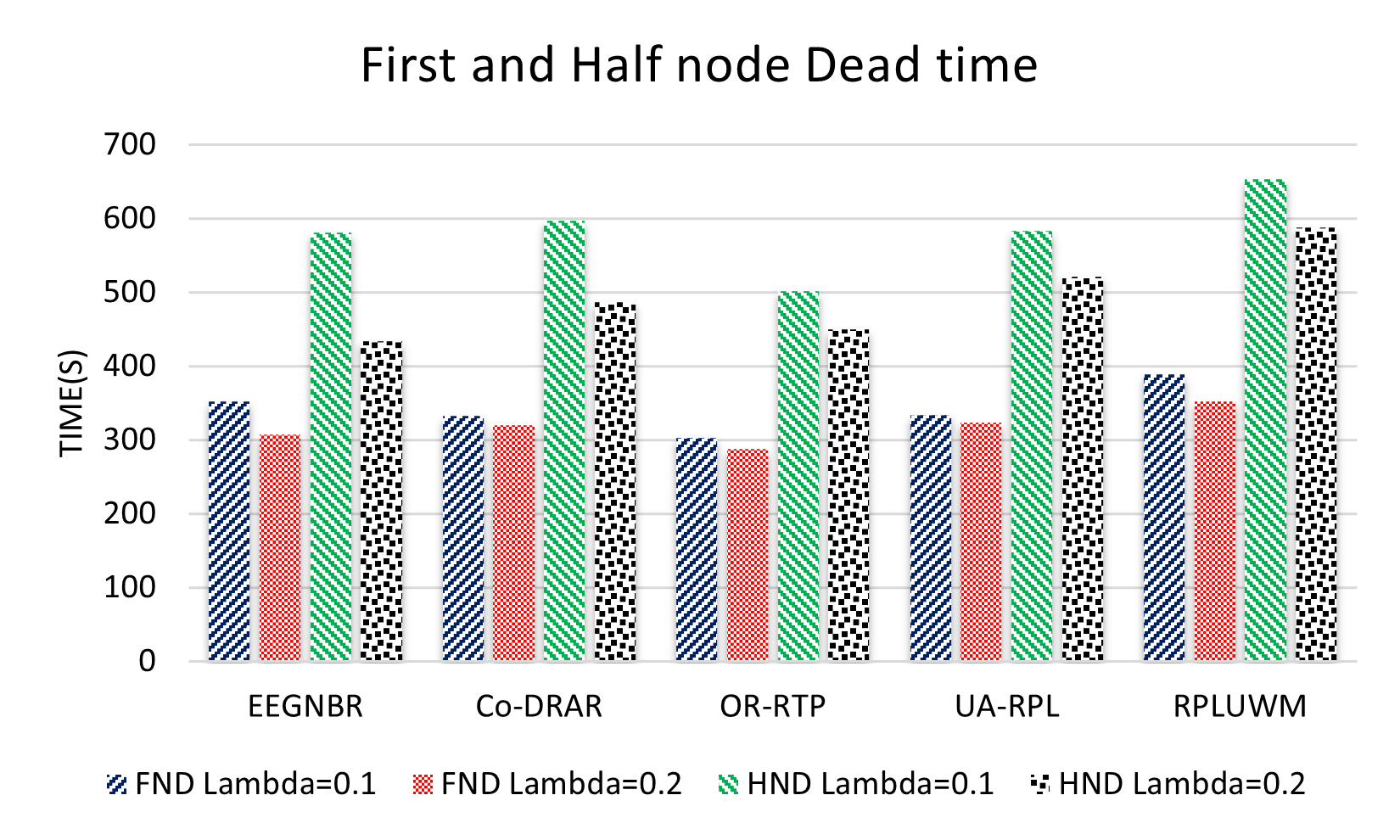}
\caption{Performance evaluation of the RPLUWM method compared with other methods, highlighting the extended lifespan of network nodes at various traffic loads.}
\label{fig:node-lifespan-dynamic}
\end{figure}

The benefits of RPLUWM were not just confined to lower traffic conditions. As the load escalated to $\lambda = 0.2$ pkt/s, the method's superiority persisted, with delays in the death time of the first node by 12.78\% for EEGNBR, 9.09\% for Co-DRAR, 18.18\% for OR-RTP, and 8.24\% for UA-RPL. Meanwhile, the advantage for the median node was maintained with a lesser reduction of 6.33\%, 5.28\%, 9.50\%, and 7.69\%, respectively. This demonstration of resilience emphasizes RPLUWM's aptitude for energy conservation and its capacity to maintain network integrity under varying node mobility and traffic load conditions. Refer to Figure~\ref{fig:node-lifespan-dynamic} for a visual representation of these comparative results.

\subsection{Average lifetime networl}
The Average Lifetime Network(ALTN) is a pivotal metric in evaluating network sustainability. This measure reflects on the equity of energy distribution within the grid and its consumption pattern after 900 seconds of simulation. A primary indicator for ALTN is the delay in the time of death of the network's first node; the longer this delay, the more effective the method is at addressing energy consumption hotspots and achieving balance. The ALTN is calculated according to Equation~\eqref{eq:ALTN}:
\begin{equation}
ALTN = \frac{\sum_{i=1}^{N-M} t_i + (M \times \wp)}{N} 
\label{eq:ALTN}
\end{equation}
where $t_i$ represents the time of death of the $i^{th}$ node, $N$ is the total number of nodes in the network, $M$ is the number of nodes surviving at the end of the simulation, and $\wp$ is the predefined lifetime of the network.

\begin{figure}[h!]
\centering
\includegraphics[width=0.6\linewidth]{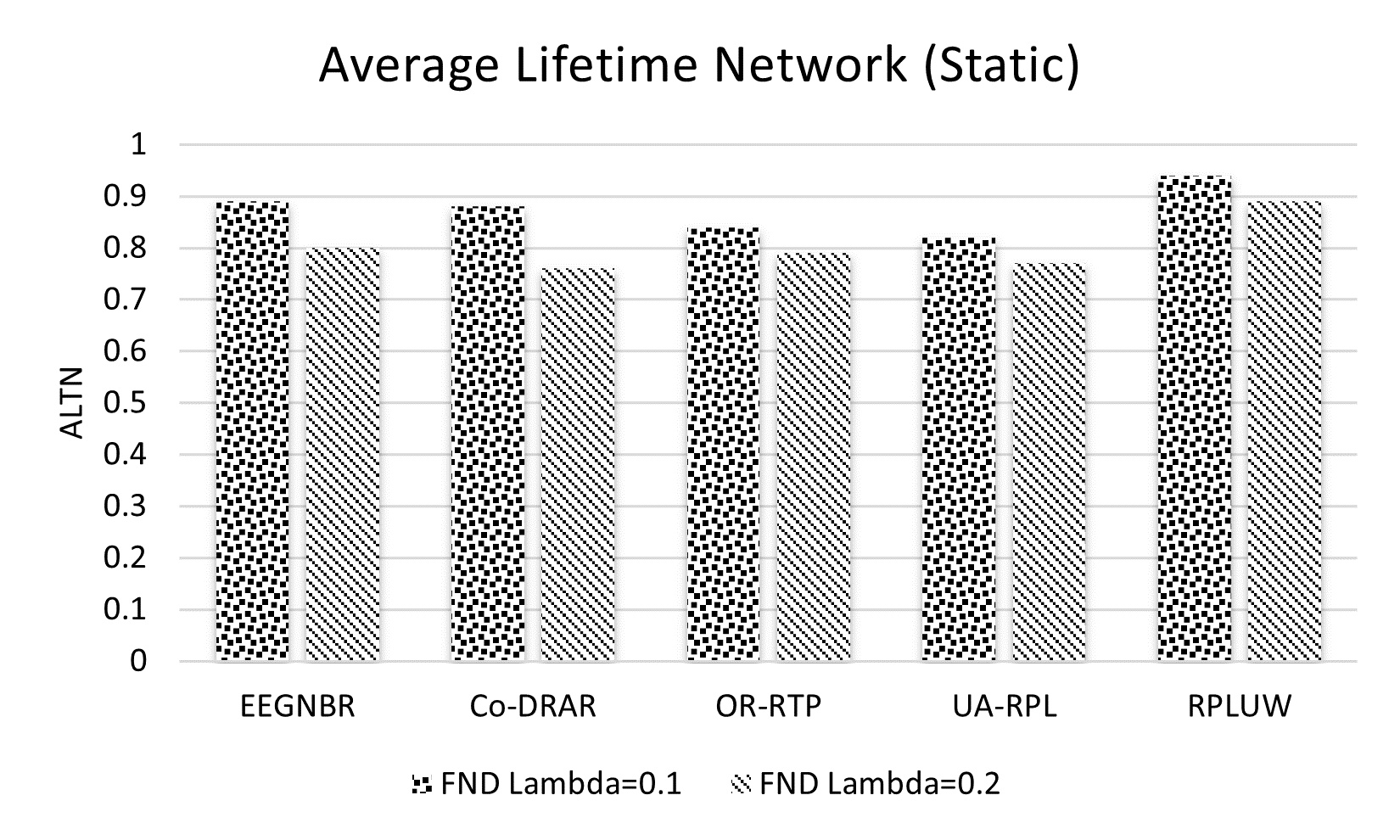}
\caption{Performance comparison showing the superior average lifetime of network nodes using RPLUW.}
\label{fig:RPLUW-ALTN}
\end{figure}

\begin{figure}[h!]
\centering
\includegraphics[width=0.6\linewidth]{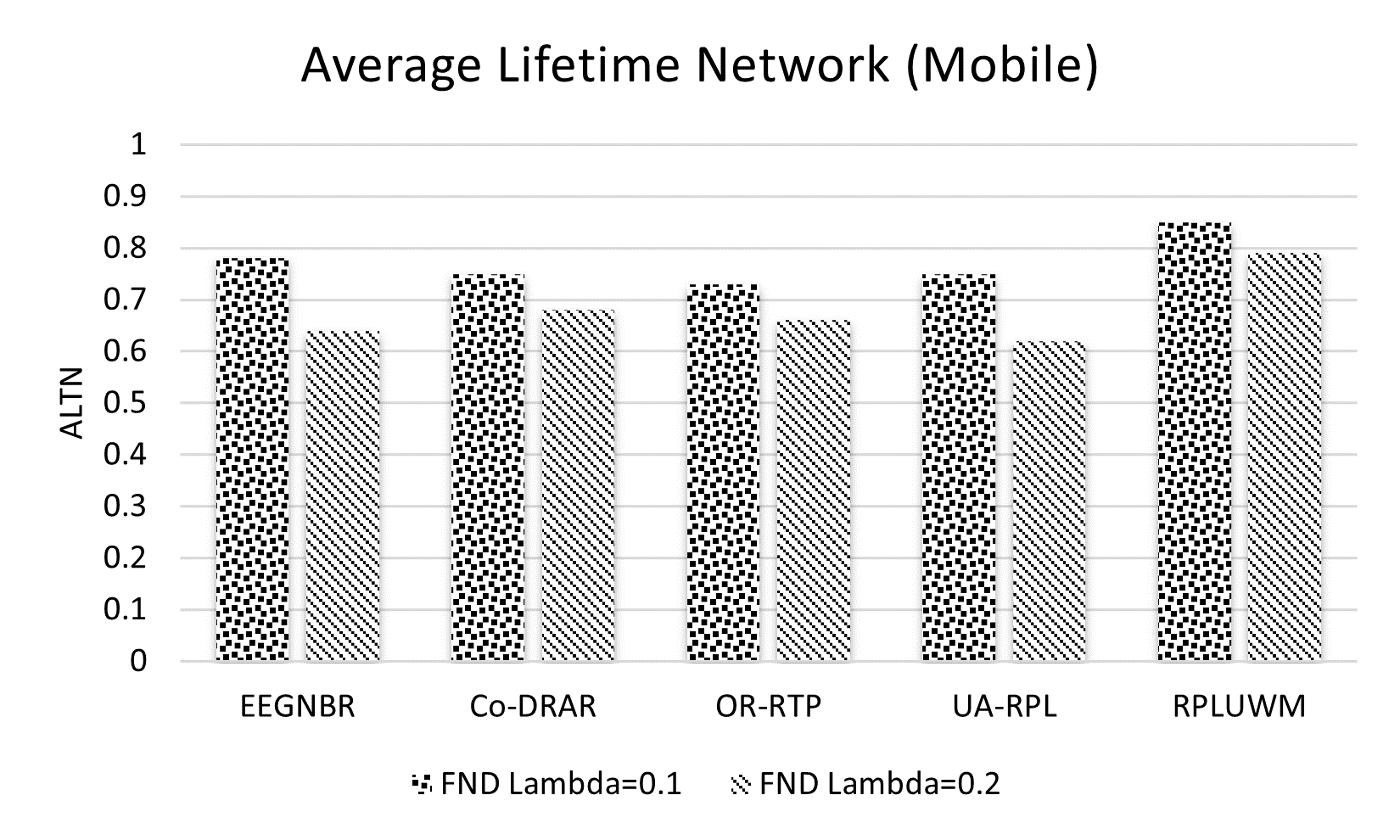}
\caption{Performance comparison showing the superior average lifetime of network nodes using RPLUWM.}
\label{fig:RPLUWM-ALTN}
\end{figure}

Figures~\ref{fig:RPLUW-ALTN} and \ref{fig:RPLUWM-ALTN} illustrate the enhancements achieved by RPLUW and RPLUWM, respectively. For a traffic input rate of $\lambda = 0.1$ to $0.2$, RPLUW has demonstrated significant improvements in the average lifetime of nodes, outperforming EEGNBR by 5.62\% to 11.25\%, Co-DRAR by 6.82\% to 17.11\%, OR-RTP by 11.90\% to 12.66\%, and UA-RPL by 14.63\% to 15.58\%. Similarly, RPLUWM has shown substantial superiority with mobile nodes, registering improvements of 8.97\% to 23.44\% over EEGNBR, 13.33\% to 16.18\% over Co-DRAR, 16.44\% to 19.70\% over OR-RTP, and 13.33\% to 27.42\% over UA-RPL. These results underscore the protocols' effectiveness in extending operational durations and maintaining robust network performance under varying traffic conditions.

\subsection{Packet Delivery Ratio}
In underwater Internet of Things networks, the Packet Delivery Ratio (PDR) is a critical performance metric, reflecting the data packet delivery success rate to the intended destination. The PDR is mathematically represented as the ratio of packets received by the destination nodes to the number of packets sent by the source nodes. Formally, the PDR can be expressed as:
\begin{equation}
\label{eq:pdr_formula}
\text{PDR} = \frac{\text{Number of packets successfully received}}{\text{Total number of packets sent}} \times 100
\end{equation}

\begin{figure}[htbp]
\centering
\includegraphics[width=0.6\linewidth]{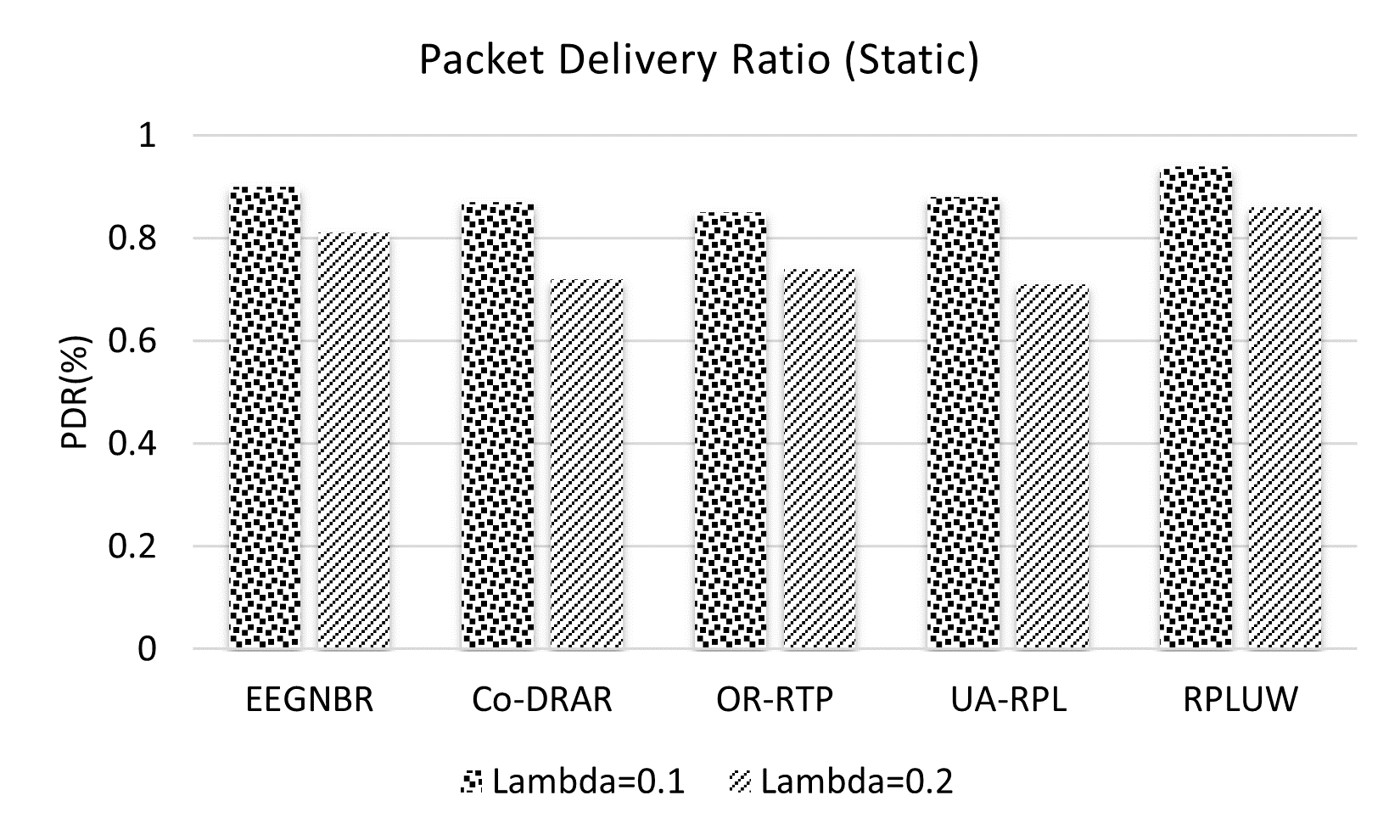}
\caption{The PDR results for static node configurations, showcasing the performance of RPLUW in comparison to traditional protocols.}
\label{fig:pdr_static}
\end{figure}

As defined in Equation (\ref{eq:pdr_formula}), the PDR for the proposed RPLUW and RPLUWM protocols was meticulously evaluated under diverse conditions characterized by static and mobile nodes. The intelligent multi-attribute decision-making approach, which synthesizes parameters such as node depth, signal strength, energy reserves, and latency, has enabled RPLUW to exhibit remarkable improvements in packet delivery. Specifically, in a static node environment, RPLUW achieved a PDR enhancement of approximately 4.44\% at $\lambda = 0.1$ pkt/s and 17.65\% at $\lambda = 0.2$ pkt/s, in comparison to existing protocols like EEGNBR, Co-DRAR, OR-RTP, and UA-RPL. These advancements are vividly depicted in Figure \ref{fig:pdr_static}.

\begin{figure}[htbp]
\centering
\includegraphics[width=0.6\linewidth]{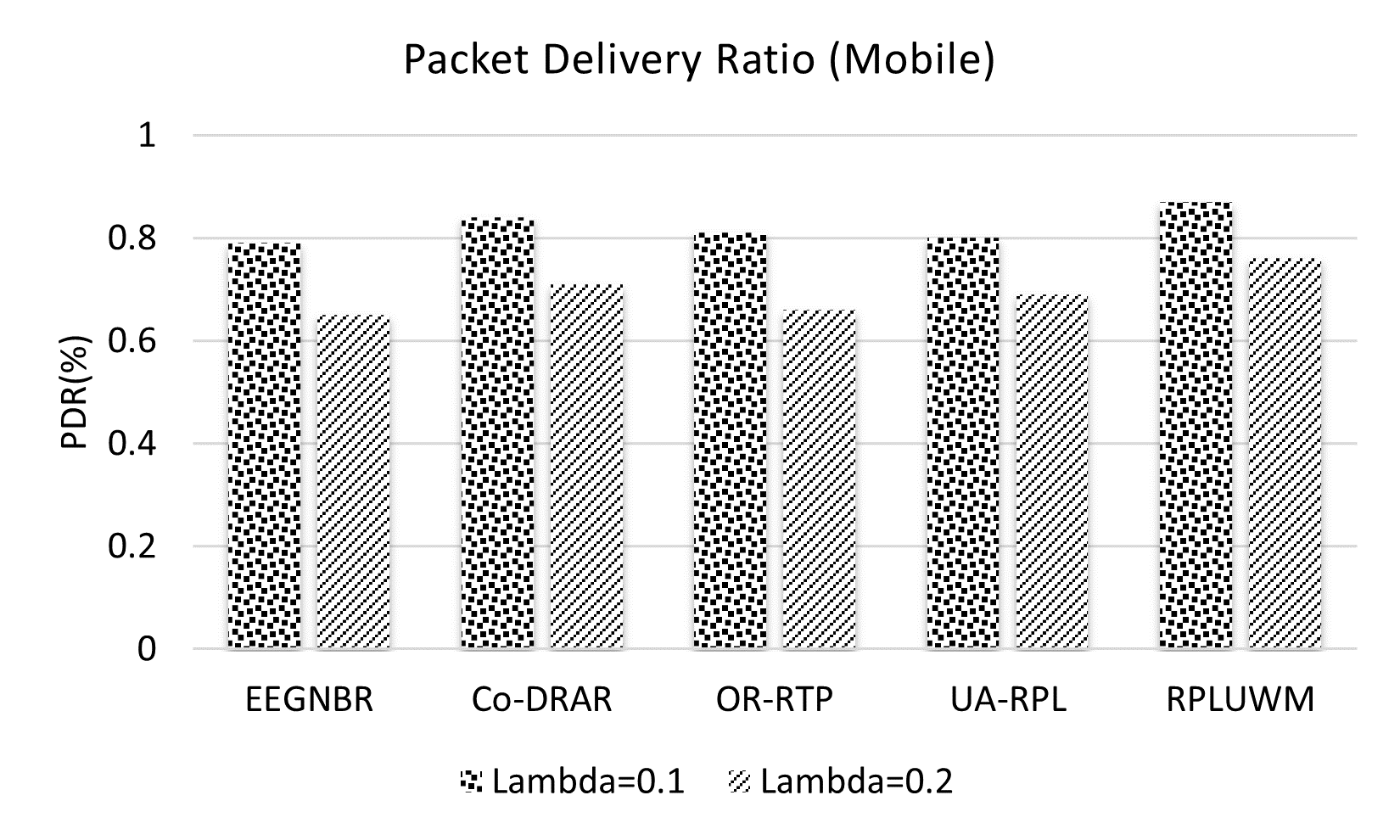}
\caption{The PDR results for mobile node configurations, illustrating the efficacy of RPLUWM under dynamic conditions.}
\label{fig:pdr_mobile}
\end{figure}

For mobile nodes, the adaptability of RPLUWM to the dynamic underwater environment facilitated PDR improvements of 14.29\% and 16.67\% for traffic rates of $\lambda = 0.1$ pkt/s and $\lambda = 0.2$ pkt/s, respectively. Despite the underwater communication challenges, the simulation results showcased in Figure \ref{fig:pdr_mobile} substantiate this capability to maintain high packet delivery rates. The findings collectively underscore the superiority of the proposed methods, solidifying the efficacy of the multi-attribute decision-making framework in enhancing the reliability of underwater sensor network communications.

\subsection{Average End-to-End Delay}
The simulation results for the RPLUW protocol, and its mobile version RPLUWM, under two different traffic rates—$\lambda=0.1$ and $\lambda=0.2$ packets per second—demonstrate significant improvements in end-to-end delay compared to other established protocols. At a $\lambda$ of 0.1, the RPLUW protocol exhibited remarkable reductions in delay; EEGNBR, Co-DRAR, and OR-RTP experienced approximately 166.95\%, 31.78\%, and 13.56\% higher delays, respectively, while UA-RPL showed a minimal but notable 3.81\% lower delay. When the traffic rate increased to $\lambda=0.2$, RPLUW's performance continued to outpace the competition, with EEGNBR, Co-DRAR, and OR-RTP delays being higher by 184.92\%, 48.02\%, and 18.25\%, respectively, and UA-RPL presenting a marginally improved delay of 1.19\%.

For the RPLUWM protocol tailored to mobile scenarios, the efficiency becomes even more pronounced. With $\lambda=0.1$, delays in EEGNBR, Co-DRAR, OR-RTP, and UA-RPL were 128.71\%, 45.50\%, 54.01\%, and 40.63\% higher, respectively. At a $\lambda$ of 0.2, RPLUWM showcased a stellar performance increase with reductions in delay compared to EEGNBR, Co-DRAR, OR-RTP, and UA-RPL by 140.86\%, 75.91\%, 93.12\%, and 54.41\%, respectively.

\begin{figure}[h!]
\centering
\includegraphics[width=0.6\linewidth]{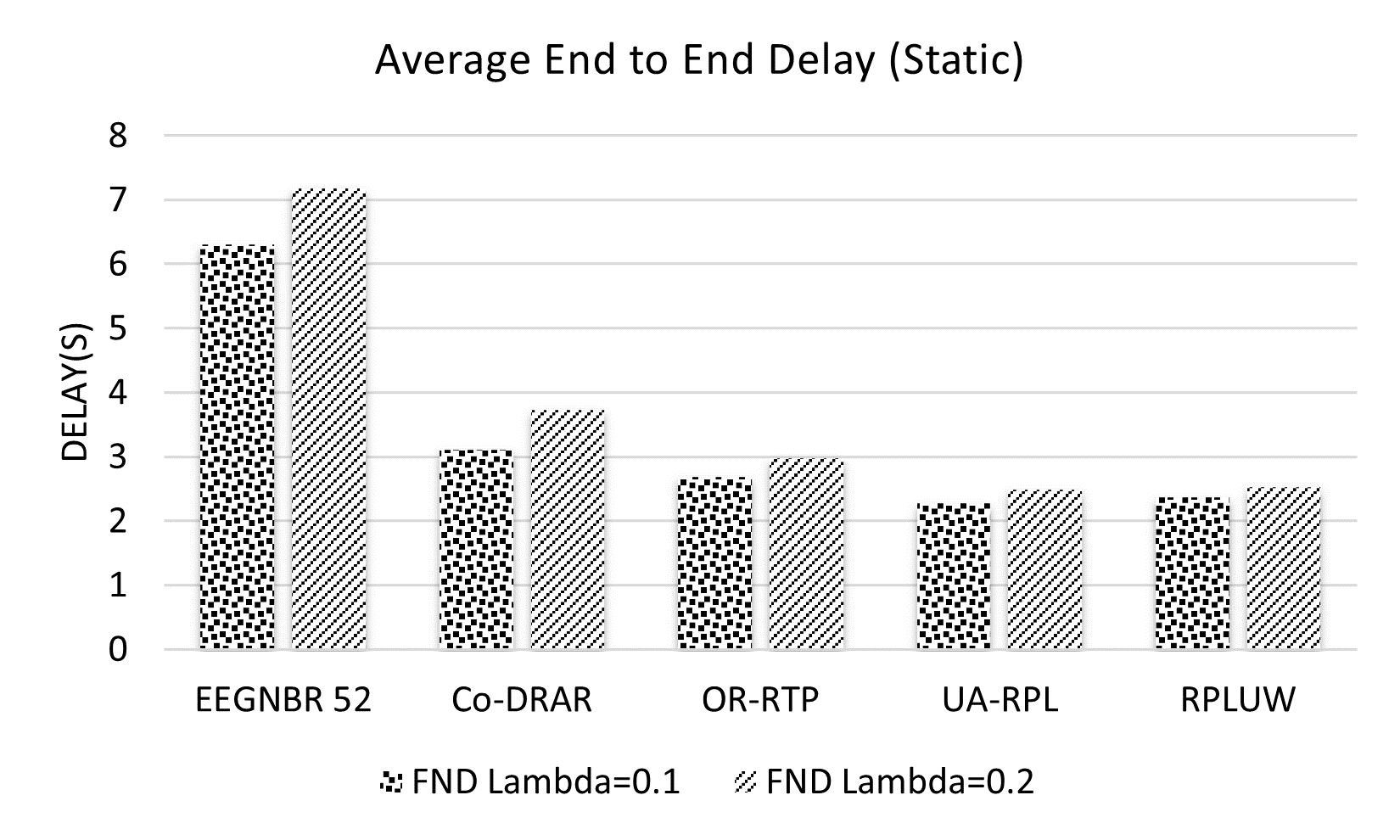}
\caption{The superior performance of the RPLUW protocol in a fixed node environment showcases significant delay reductions.}
\label{fig:RPLUW_fixed-end2end-delay}
\end{figure}

\begin{figure}[h!]
\centering
\includegraphics[width=0.6\linewidth]{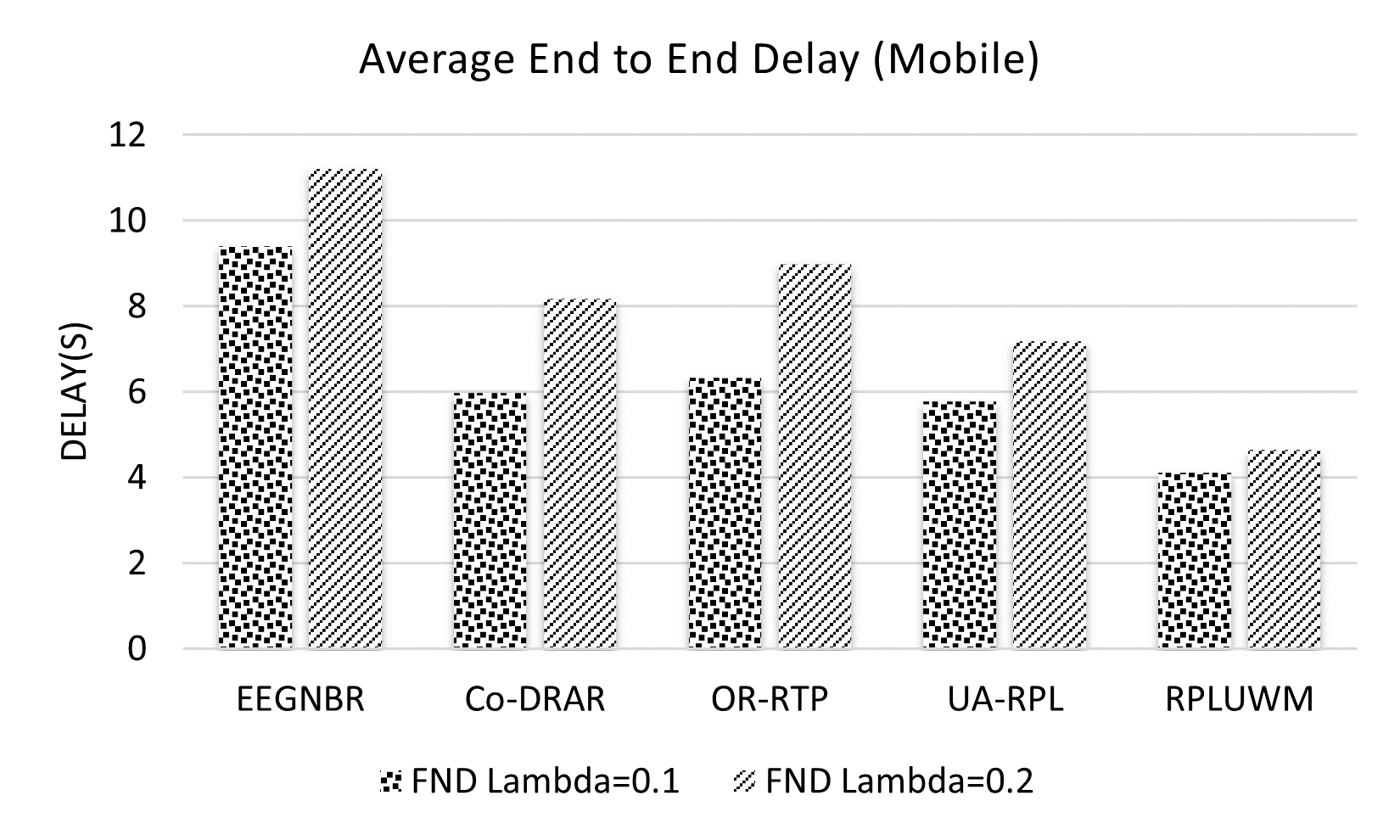}
\caption{The superior performance of the RPLUWM protocol in a mobile node environment demonstrates substantial improvements in end-to-end delay.}
\label{fig:RPLUW_mobile-end2end-delay}
\end{figure}

End-to-end latency is a vital metric for the efficiency of underwater sensor networks, reflecting network performance in environmental monitoring tasks. The RPLUW and RPLUWM protocols' decreased jitter and latency signify robust environmental monitoring capabilities. Specifically, RPLUW with stationary nodes improved end-to-end latency by a minimum of 3.81\% and a maximum of 166.95\% across various protocols. In contrast, the mobile adaptation RPLUWM secured a minimum improvement of 40.63\% and a maximum of 140.86\%. These results underscore the efficacy of the proposed protocols' scheduling, multi-route routing, and decision-making systems within the network. This enhancement is attributed to the protocols' consideration of critical factors in establishing and maintaining link longevity between nodes. Figures \ref{fig:RPLUW_fixed-end2end-delay} and \ref{fig:RPLUW_mobile-end2end-delay} depict fixed and mobile node environments, respectively.

\subsection{Network convergence time test}

\begin{figure}[htbp]
\centering
\includegraphics[width=0.6\linewidth]{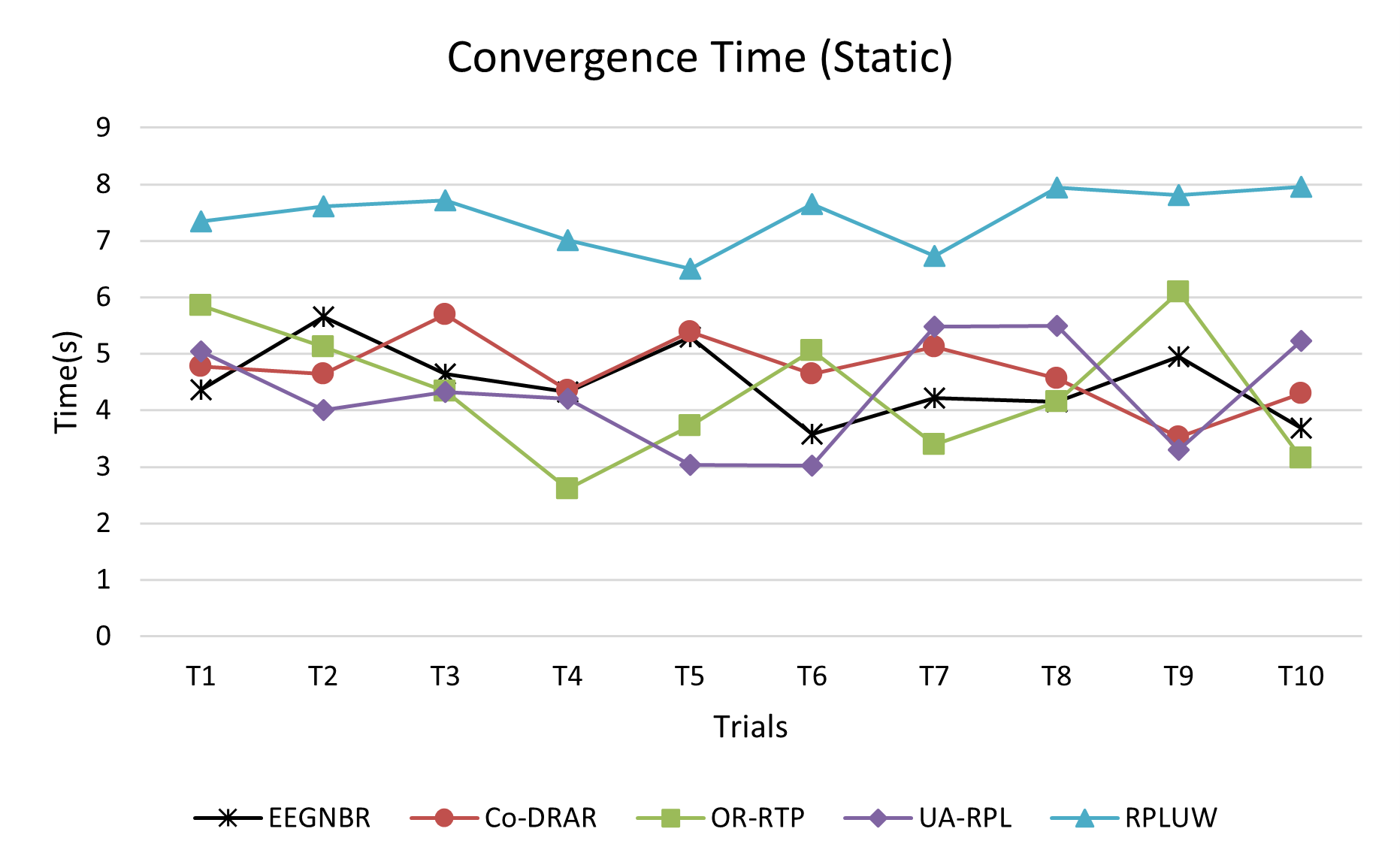}
\caption{Network topology convergence in Static Nodes.}
\label{fig:convergence_static}
\end{figure}

\begin{figure}[htbp]
\centering
\includegraphics[width=0.6\linewidth]{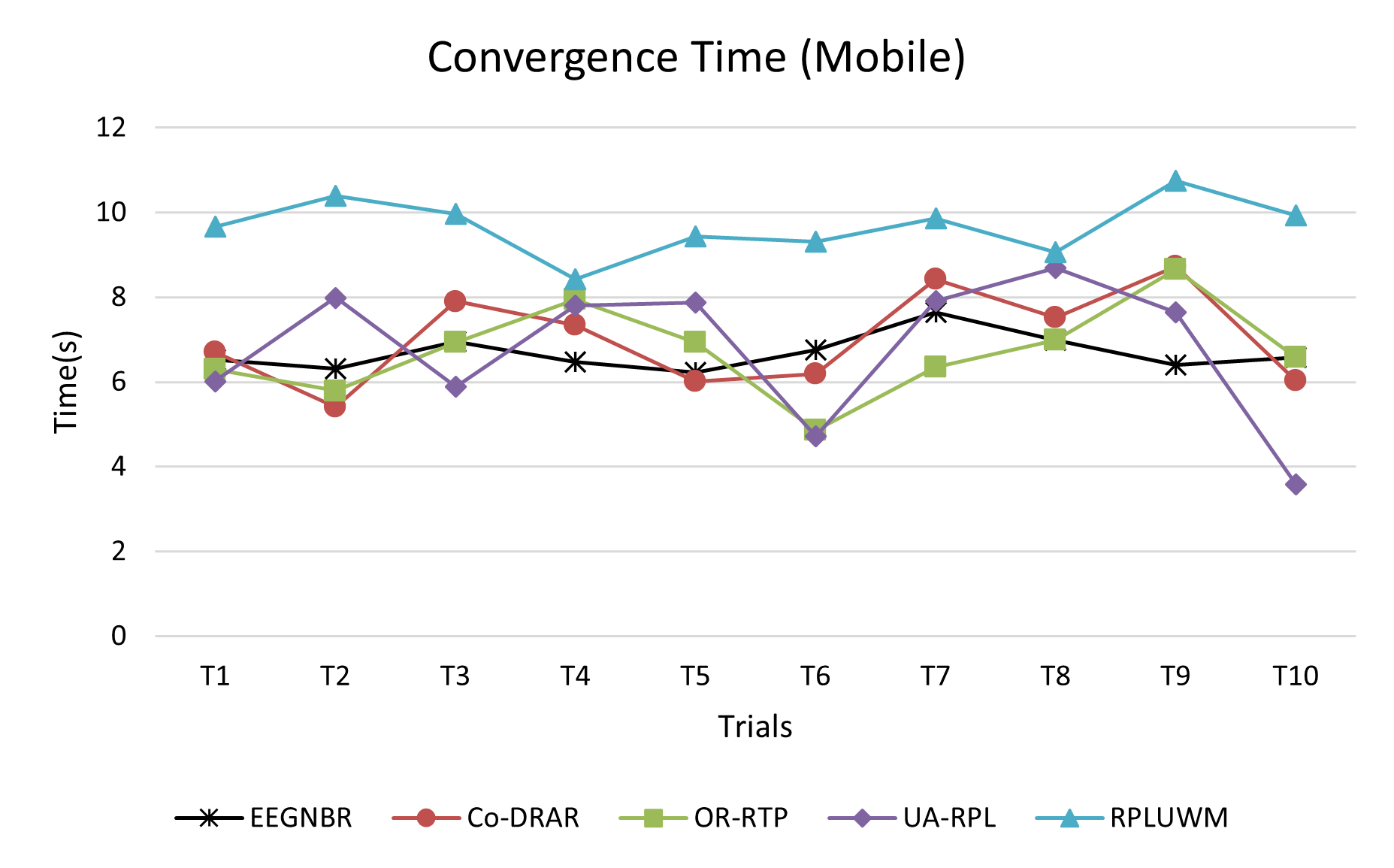}
\caption{Network topology convergence in Mobile Nodes.}
\label{fig:convergence_mobile}
\end{figure}

In the realm of underwater sensor networks, achieving optimal performance across various parameters remains a complex trade-off. Comparative analysis of different protocols in terms of energy consumption, network lifespan, packet delivery ratio, delay, and the time to first and half-node deaths reveals that our proposed protocols, RPLUW and RPLUWM, excel in most performance metrics (see Figures \ref{fig:convergence_static} and \ref{fig:convergence_mobile}). They demonstrate enhanced energy efficiency, increased average network lifetime, and superior packet delivery rates with lower overall delays. These protocols also show a delayed occurrence of the first and half-node deaths, indicating robustness in network functionality over time. However, these advantages come at the cost of a longer network convergence time. This trade-off emphasizes that while RPLUW and RPLUWM protocols offer significant improvements in sustaining network performance and reliability, they require a larger initial investment to establish network routes. This characteristic is a strategic choice that prioritizes long-term operational benefits over immediate network readiness, marking the protocols as particularly suited for applications where network durability and resilience are paramount and initial setup time can be afforded.

\newpage
\section{CONCLUSION} \label{sec6}
As hardware technologies advance and telecommunications networks become increasingly efficient, implementing IoT networks to monitor physical parameters has become more streamlined and possible. In light of the diverse applications of the Internet of Things within aquatic environments, there is a pressing need for efficient platforms that can extend operational life and minimize latency, particularly in applications where timing is crucial. Our research introduced a suite of pivotal parameters that contribute to establishing, upkeep, and restoring network topology through a multi-criteria decision-making framework. Additionally, we addressed the complexities of node mobility within the network by integrating innovative trickle timers and refined algorithms for neighbour discovery. Adopting a multi-route approach has been instrumental in diminishing node traffic and achieving a harmonious load distribution. The employment of decision-making systems requiring minimal computational resources has notably enhanced the topology management and network graph restoration processes. Our simulation results, juxtaposed with contemporary methods, demonstrate the superior functionality and effectiveness of our proposed RPLUW and RPLUWM protocols. Achievements such as prolonged network longevity, improved delivery rates, and reduced end-to-end latency underscore the successes of this study. Looking forward, we aim to tackle the intricacies of data aggregation in underwater IoT networks by proposing a model that focuses on energy optimization and time efficiency, rooted in the foundational RPLUW and RPLUWM strategies and bolstered by learning algorithms. The remarkable prolongation of network node lifespans via RPLUWM signals its promise for sustained operation in underwater monitoring endeavours, vital for environmental research, resource prospecting, and security activities. This innovation marks a significant leap in underwater communications, presenting a resilient framework that plays a transformative role in the future of marine exploration and data acquisition.

\section*{Acknowledgment}
\clearpage



\section*{Author Biographies}\label{sec9}

\noindent\begin{tabular}{m{0.25\textwidth} m{0.7\textwidth}}
    \adjustbox{valign=t}{\includegraphics[width=0.25\textwidth]{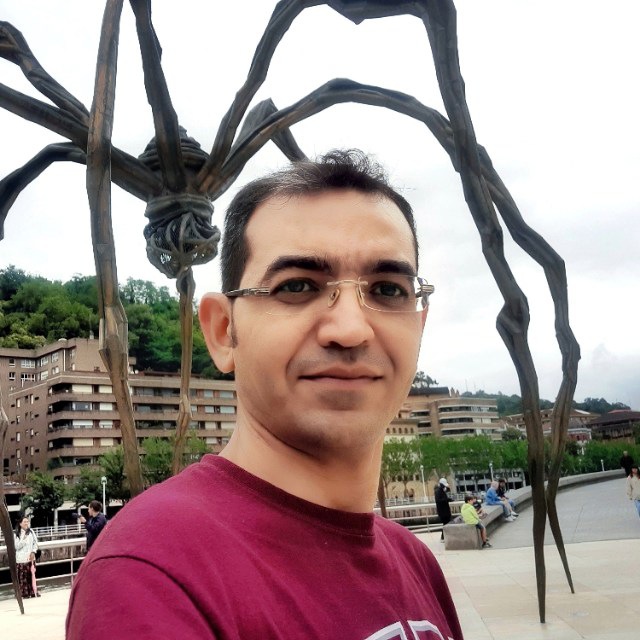}} &
    \textbf{MOHAMMADHOSSEIN HOMAEI (M'19)} was born in Hamedan, Iran. He obtained his B.Sc. in Information Technology (Networking) from the University of Applied Science and Technology, Hamedan, Iran, in 2014 and his M.Sc. from Islamic Azad University, Malayer, Iran, in 2017. He is pursuing his Ph.D. at Universidad de Extremadura, Spain, where his prolific research has amassed over 100 citations.
    
    Since December 2019, Mr. Homaei has been affiliated with Óbuda University, Hungary, as a Visiting Researcher delving into the Internet of Things and Big Data. His tenure at Óbuda University seamlessly extended into a research collaboration with J. Selye University, Slovakia, focusing on Cybersecurity from January 2020. His research voyage led him to the National Yunlin University of Science and Technology, Taiwan, where he was a Scientific Researcher exploring IoT and Open-AI from January to September 2021. His latest role was at the Universidade da Beira Interior, Portugal, in the Assisted Living Computing and Telecommunications Laboratory (ALLab), from June 2023 to January 2024, where he engaged in cutting-edge projects on digital twins and machine learning. He is the author of ten scholarly articles and holds three patents, highlighting his diverse research interests in Digital Twins, Cybersecurity, Wireless Communications, and IoT. 
    
    An active IEEE member, Mr. Homaei has carved a niche for himself with notable contributions to Digitalization, the Industrial Internet of Things (IIoT), Information Security Management, and Environmental Monitoring. His substantial work continues to influence the technological and cybersecurity landscape profoundly. 
\end{tabular}

\newpage

\section{Appendixes}\label{sec7}

\begin{itemize}
    \item \textbf{Shallow Water Energy Consumption Model:} This model assesses the energy consumption in shallow waters. It is based on a linear distribution of sensor nodes featuring \(N + 1\) nodes where the distance between consecutive nodes is denoted as \(d\). The calculations in this model address the transmission of packets containing \(B\) bits to the sink node, using either a single-hop or a multi-hop process involving relay nodes. The model adopts a linear chain configuration of nodes, which is considered a worst-case scenario for analysis. As illustrated in Figure 3, the propagation of sound signals in shallow waters is modeled cylindrically, necessitating the application of cylindrical spatial geometry for accurate calculations.

\begin{figure}[htbp]
\centering
\includegraphics[width=0.25\linewidth]{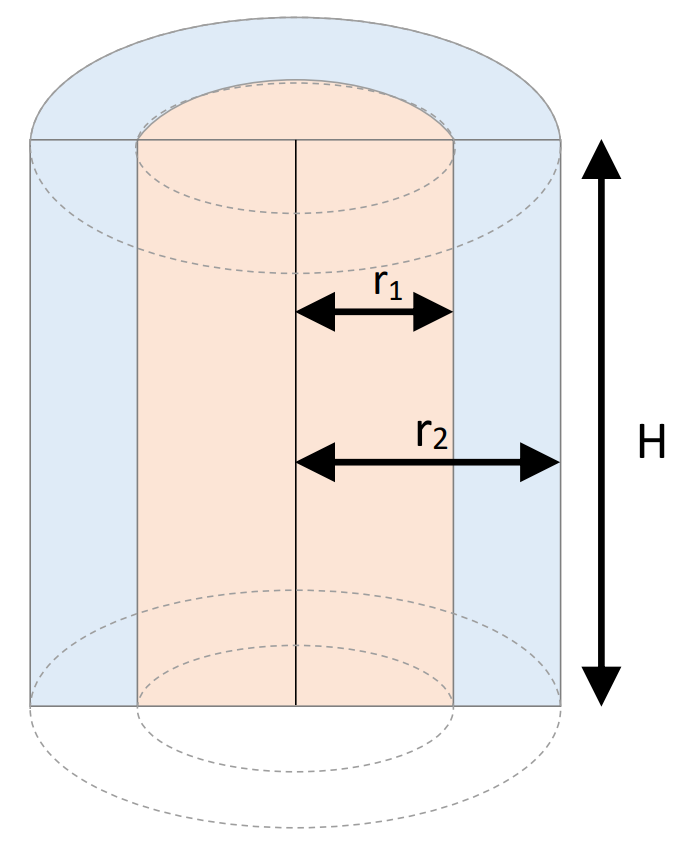}
\caption{Cylindrical environment around each underwater sensor node.}
\label{fig:cylindrical_environment}
\end{figure}

As illustrated in Figure \ref{fig:cylindrical_environment}, we utilize the subsequent equations to calculate the transmission power at two distinct points, \( r_1 \) and \( r_2 \):

\begin{equation}
P = 2\pi r_1 H_1 \label{eq:power_r1}
\end{equation}

\begin{equation}
P = 2\pi r_2 H_2 \label{eq:power_r2}
\end{equation}

The variable \( H \) is defined as the height of the cylinder, equivalent to the depth of the node. The rate at which transmission diminishes between \( r_1 \) and \( r_2 \) is determined by Equation \ref{eq:transmission_drop}:

\begin{equation}
TL = 100\log{\frac{r_1}{r_2}} \label{eq:transmission_drop}
\end{equation}

Equations \ref{eq:power_r1} and \ref{eq:power_r2} describe the transmission power related to the radial distances \( r_1 \) and \( r_2 \) from the central axis of the cylindrical underwater environment surrounding a sensor node, as depicted in Figure \ref{fig:cylindrical_environment}.

Now consider that a node located at a distance of \( Nd \) from the sink node needs to send \( K \) packets. The required power level and energy consumption during its transfer are calculated through Equations \ref{eq:power} and \ref{eq:energy}:

\begin{equation}
P = 2\pi d H_1 \label{eq:power}
\end{equation}

\begin{equation}
E = N P_{T_{x}} K \label{eq:energy}
\end{equation}

The parameter \( d \) is the distance between two nodes, \( N \) indicates the number of steps to the Sink, \( T_{tx} \) is the transmission time of a packet. When each node is in the process of transferring \( m \) packets and this transfer is a multi-hop relay mechanism, the energy consumption is equal to Equation~\ref{eq:multi_hop_energy2}:

\begin{align}
E_{\text{total}} &= NP_{T}t_{x}m + (N - 1)P_{T}t_{x}m + \ldots + P_{T}t_{x}m \nonumber \\
&= \frac{N(N+1)P_{T}t_{x}m}{2} \label{eq:multi_hop_energy2}
\end{align}

However, if the sensor node wants to interact with a single-hop connection and directly with the sink, then the power consumption of the node will be obtained from Equation~\ref{eq:single_hop_power2}:

\begin{equation}
P = 2\pi r_{1}H_{1} \label{eq:single_hop_power2}
\end{equation}

In this relation \( r_{1} \) is the distance between each node and the sink. Total energy consumption in this scenario uses Equation~\ref{eq:total_energy}:

\begin{align}
E_{\text{total}} &= P_{N}d T_{tx}m + P_{(N-1)d} T_{tx}m + \ldots + P_{d} T_{tx}m \nonumber \\
&= mT \sum_{i=1}^{N} P(i \cdot d) \label{eq:total_energy}
\end{align}

    \item \textbf{Deep Water Energy Consumption Model:} Unlike shallow waters, the propagation of sound signals in deep water is spherical. In a network scenario, as in the case of shallow water, the power P is generated at the source and propagated in all directions as a sphere, as shown in Figure \ref{fig:Propagation-model-deep-water}.

\begin{figure}[htbp]
\centering
\includegraphics[width=0.2\linewidth]{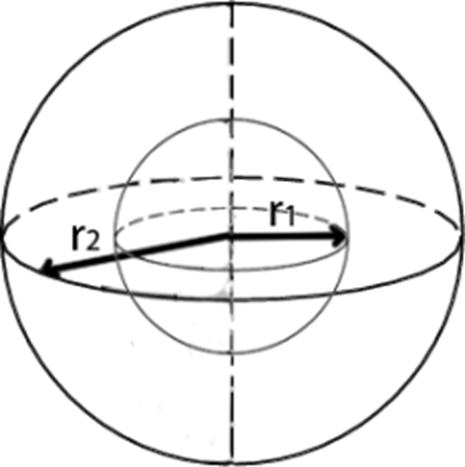}
\caption{Signal propagation model in deep water.}
\label{fig:Propagation-model-deep-water}
\end{figure}
 
\end{itemize}

Thus, if we consider the points \( r_1 \) and \( r_2 \) as shallow water, then the signal strength is obtained from Equation~\ref{eq:signal_strength}:

\begin{equation}
P = 4\pi r_1^2 I_1 = 4\pi r_2^2 I_2 \label{eq:signal_strength}
\end{equation}

The transfer loss rate between points \( r_1 \) and \( r_2 \) is also calculated from Equation~\ref{eq:transfer_loss_rate}:

\begin{equation}
TL = 10 \log \frac{I_1}{I_2} = 20 \log r_2 \label{eq:transfer_loss_rate}
\end{equation}

The power and energy consumption levels for a network topology when the node is at a distance \( Nd \) from the sink and wants to transmit data in several hops will be calculated from Equations~\ref{eq:multi_hop_power} and \ref{eq:multi_hop_energy}:

\begin{equation}
P = 4\pi d^2 I_1 \label{eq:multi_hop_power}
\end{equation}

\begin{equation}
E = NP_{T_{x}} m \label{eq:multi_hop_energy}
\end{equation}

Whenever in the linear distribution, each node needs to transfer \( K \) packets, the energy consumption in the multi-hop relay scenario is obtained by Equation~\ref{eq:multi_hop_energy}. If the node wants to transmit in a single hop, the received signal must be generated with the power resulting from Equation~\ref{eq:single_hop_power}:

\begin{equation}
P = 4\pi r_1^2 I_1 \label{eq:single_hop_power}
\end{equation}

\( r_1 \) is the distance between the node and the sink. However, unlike in the case of shallow water, the general drop or rate of loss of a data transmission is a combination of spherical propagation, signal attenuation, and environmental incompatibility. Therefore, the amount of signal attenuation is calculated from Equation~\ref{eq:signal_attenuation}:

\begin{equation}
TL = 20 \log r + \alpha r \times 10^{-3} + A \label{eq:signal_attenuation}
\end{equation}

\end{document}